\documentclass[11pt,a4paper]{article}
\pdfoutput=1
\usepackage{jcappub}
\usepackage{slashed}
\usepackage{bm} 
\usepackage{epsf}
\usepackage{graphicx,epsfig}
\usepackage{latexsym,amssymb,amsmath,float,url}
\usepackage{bbold}
\usepackage{latexsym}
\usepackage{soul}
\usepackage{dsfont}
\usepackage{natbib}

\input{epsf}

\newcommand{\nocontentsline}[3]{}
\newcommand{\tocless}[2]{\bgroup\let\addcontentsline=\nocontentsline#1{#2}\egroup}

\title{The role of CP violating scatterings in baryogenesis --- case study of the neutron portal}

\author[a]{Iason Baldes,}
\author[a]{Nicole F. Bell,}
\author[a]{Alexander Millar,}
\author[b]{Kalliopi Petraki}
\author[a]{and Raymond R. Volkas.}
\affiliation[a]{ARC Centre of Excellence for Particle Physics at the Terascale, \\
School of Physics, The University of Melbourne, Victoria 3010, Australia}
\affiliation[b]{Nikhef, Science Park 105, 1098 XG Amsterdam, The Netherlands}

\emailAdd{i.baldes@student.unimelb.edu.au}
\emailAdd{n.bell@unimelb.edu.au}
\emailAdd{a.millar@student.unimelb.edu.au}
\emailAdd{kpetraki@nikhef.nl}
\emailAdd{raymondv@unimelb.edu.au}

\date{1 October, 2014}

\abstract{Many baryogenesis scenarios invoke the charge parity (CP) violating out-of-equilibrium decay of a heavy particle in order to explain the baryon asymmetry. Such scenarios will in general also allow CP violating scatterings. We study the effect of these CP violating scatterings on the final asymmetry in a neutron portal scenario. We solve the Boltzmann equations governing the evolution of the baryon number numerically and show that the CP violating scatterings play a dominant role in a significant portion of the parameter space.
}

\date{today}

\begin{document}
\maketitle
\section{Introduction}
A baryon symmetric universe would result in a baryon density around ten orders of magnitude smaller than observed~\cite{PhysRevLett.17.712}. Deductions of the baryon asymmetry from the cosmic microwave background give a baryon-to-entropy density ratio of~\cite{Hinshaw:2012aka,Ade:2013zuv},
	\begin{equation}
	Y_{B}=(0.86 \pm 0.01)\times10^{-10},
	\end{equation}
which is in agreement with the baryon asymmetry required to match observed primordial element abundances: $0.4\times10^{-10} \lesssim Y_{B} \lesssim 0.9\times10^{-10}$~\cite{Copi:1994ev}. The required baryon asymmetry can be created dynamically by scenarios satisfying the Sakharov conditions~\cite{Sakharov:1967dj}. One way of producing an asymmetry is through the out-of-equilibrium and CP violating decays of a heavy particle such as in leptogenesis or grand unified theory baryogenesis~\cite{Weinberg:1979bt,Kolb:1979qa,PhysRevLett.45.2074,Fukugita:1986hr}. 

Scatterings of particles in $2 \leftrightarrow 2$ type processes can also violate CP and potentially produce asymmetries. In such scenarios, the out-of-equilibrium condition can be satisfied in two ways. The first option is to have a visible and a hidden sector at different temperatures \cite{Bento:2001rc,Hook:2011tk,Unwin:2014poa}; the second is to have one or more of the particles freeze out, such as  the annihilating DM particles in WIMPy baryogenesis scenarios~\cite{Cui:2011ab,Bernal:2012gv,Bernal:2013bga,Kumar:2013uca,Racker:2014uga}. The general principles behind such a freeze out scenario have been investigated in ref.~\cite{Baldes:2014gca}. CP violation in $2 \leftrightarrow 2$ processes could also provide a mechanism for producing asymmetric dark matter~\cite{Nussinov:1985xr,Davoudiasl:2012uw,Petraki:2013wwa,Zurek:2013wia} in both freeze-in~\cite{Bento:2001rc,Hook:2011tk,Unwin:2014poa} and freeze-out type scenarios~\cite{Farrar:2005zd,Baldes:2014gca}. 

The presence of CP violation in scatterings is, however, not confined to models specifically constructed to exploit this feature: it can also appear in baryogenesis-via-decay type scenarios. For example, scatterings of the heavy Majorana neutrinos in leptogenesis also violate CP~\cite{Pilaftsis:2003gt,Pilaftsis:2005rv,Nardi:2007jp,Davidson:2008bu,Fong:2010bh}. Numerical calculations show including CP violating scatterings in leptogenesis has a large effect on the baryon asymmetry at high temperature but only a negligible effect at low temperature~\cite{Pilaftsis:2003gt}. The purpose of this paper is to study the effects of CP violating scatterings on the generation of the baryon asymmetry more generally. We choose to study a neutron portal model in which the parameters are not restricted by having to explain low energy neutrino data. Baryogenesis via CP violating decays in the neutron portal have been studied previously~\cite{Cheung:2013hza}. CP violating scatterings in a neutron portal model were discussed in ref.~\cite{Farrar:2005zd}; however, the unitarity constraint was not properly taken into account. We will extend the analysis to include a full numerical treatment of the Boltzmann equations including CP violating scatterings. Such effects are not constrained to this particular model and will play a role in baryogenesis scenarios more generally.

The layout of the paper is as follows. In section \ref{sec:notation} we introduce our notation. In section~\ref{sec:model} we introduce the neutron portal Lagrangian and discuss the relevant decays and scatterings. In section~\ref{sec:boltzmanneq} we write down the Boltzmann equations for the evolution of the baryon asymmetry and show example numerical solutions. Constraints on the model are discussed in section~\ref{sec:constraints}. Possible UV completions are discussed in section~\ref{sec:uvcomplete}.

\section{Notation}
\label{sec:notation}
 In an expanding universe with no collisions, a particle number density, $n_{X}$, drops as $R^{-3}$ where $R$ is the scale factor. Taking into account collisions the density evolves as:
	\begin{equation}
	\frac{dn_{X}}{dt}+3Hn_{X}=C(X),
	\end{equation}
where $H=\dot{R}/R$ is the Hubble expansion rate and $C(X)$ is the collision term, i.e. the rate of change of $n_{X}$ due to interactions with other particles. Let us imagine now such an interaction from state $\alpha$ to state $\beta$ and its reverse process which changes $X$ number by one unit. The collision term for this transition can be written as \cite{earlyuniverse}:
	\begin{align}
	C_{\alpha \beta}(X) = \int...\int   d\Pi_{\alpha 1}...d\Pi_{\alpha n}d\Pi_{\beta 1}...d\Pi_{\beta m}\delta^{4}\left(\sum p_{i} - \sum p_{j}\right)(2\pi)^{4} \nonumber \\
		 \times\Big\{f_{\beta 1}...f_{\beta m}|\mathcal{M}(\beta \to \alpha)|^{2} -f_{\alpha 1}...f_{\alpha n}|\mathcal{M}(\alpha\to \beta)|^{2}\Big\}, \label{eq:maxboltcollision}
	\end{align}
where $f_{\psi}=\mathrm{Exp}[(\mu_{\psi}-E_{\psi})/T]$ is the phase space density of species $\psi$ with chemical potential $\mu_{\psi}$ at energy $E_{\psi}$ and temperature $T$,
	\begin{equation} 	
	d\Pi_{\psi}=\frac{g_{\psi}d^{3}p_{\psi}}{2E_{\psi}(2\pi)^{3}}
	\end{equation}
is the normalised volume element of the three momentum and $g_{\psi}$ counts the degrees of freedom of $\psi$. As an approximation we have used Maxwell-Boltzmann statistics and so have dropped the stimulated emission factors and Pauli blocking terms appropriate for the full quantum statistics. We assume kinetic equilibrium, i.e. a common temperature $T$, throughout. We denote the equilibrium reaction rate density for a process $\alpha \to \beta$ as:
	\begin{align}
	W(\alpha \to \beta)  \equiv & \int...\int   d\Pi_{\alpha 1}...d\Pi_{\alpha n}d\Pi_{\beta 1}...d\Pi_{\beta m}\delta^{4}\left(\sum p_{i} - \sum p_{j}\right)(2\pi)^{4} f_{\alpha 1}^{eq}...f_{\alpha n}^{eq}|\mathcal{M}(\alpha\to \beta)|^{2} \nonumber \\
		   \equiv & n_{\alpha  1}^{eq}...n_{\alpha n}^{eq}\langle v \sigma(\alpha \to \beta) \rangle, 
	\label{eq:a1}
	\end{align}
where $\langle v \sigma(\alpha \to \beta) \rangle$ is the thermally averaged cross section and $f_{\alpha i}^{eq}$ ($n_{\alpha i}^{eq}$) denotes the phase space (particle number) density in the absence of a chemical potential. Using Maxwell-Boltzmann statistics:
	\begin{equation}
	n_{\alpha i}^{eq} = \frac{g_{\alpha i}M_{\alpha i}^{2}T}{2\pi^2}K_{2}\left(\frac{M_{\alpha i}}{T}\right),
	\end{equation}
where $M_{\alpha i}$ is the mass of the species and $K_{2}(x)$ is the modified Bessel function of the second kind of order two. In the Maxwell-Boltzmann approximation, the non-equilibrium rate is found using the appropriate re-weighting:
	\begin{equation}
	W^{\mathrm{neq}}(\alpha \to \beta)=\frac{ n_{\alpha 1}...n_{\alpha n} }{ n_{\alpha 1}^{eq}...n_{\alpha n}^{eq} } W(\alpha \to \beta).
	\end{equation} 
Unitarity of the S-matrix and invariance under charge parity time (CPT) gives a condition for the equilibrium rate densities \cite{Weinberg:1979bt,Kolb:1979qa}:
	\begin{equation}
	\sum_{\beta}W(\alpha \to \beta) = \sum_{\beta}W(\beta \to \alpha) = \sum_{\beta}W(\overline{\beta} \to \overline{\alpha}) = \sum_{\beta}W(\overline{\alpha} \to \overline{\beta}),
	\label{eq:eqcond}
	\end{equation}
where the sum runs over all possible final states and $\overline{\alpha}$ denotes the CP conjugate of $\alpha$. This ensures no particle number asymmetry can be generated without a departure from equilibrium \cite{Toussaint:1978br,Dolgov:1979mz,PhysRevD.19.3803}.

\section{Neutron portal}
\label{sec:model}
\subsection{Lagrangian}
Consider the interaction Lagrangian~\cite{Cheung:2013hza},
	\begin{align}
	\Delta\mathcal{L}=  &\kappa_{1ijk}\overline{X_{1L}}u_{Ri}\overline{(d_{Rj})^{c}}d_{Rk}+\kappa_{2ijk}\overline{X_{2L}}u_{Ri}\overline{(d_{Rj})^{c}}d_{Rk}+\kappa_{3ij}\overline{u_{Ri}}X_{1L}\overline{X_{2L}}u_{Rj} \nonumber \\
	& +\frac{1}{2}\kappa_{4ij}\overline{u_{Ri}}X_{1L}\overline{X_{1L}}u_{Rj}+\frac{1}{2}\kappa_{5ij}\overline{u_{Ri}}X_{2L}\overline{X_{2L}}u_{Rj}+H.c.,
	\label{eq:eftlag}
	\end{align}
where the $\kappa_{a ijk}$ are couplings with units of (mass)$^{-2}$, $u_{Ri}$ ($d_{Ri}$) is the right chiral up (down) type quark field of flavour $i$, $\Psi^{c}$ denotes the conjugate of field $\Psi$ and the $X_{\alpha}$ are Majorana fermions. We impose a global lepton number symmetry forbidding coupling of the $X_{\alpha}$ to the standard model (SM) Higgs and lepton doublets. Spinor indices are contracted between the first two and last two fermion fields in each of the terms. We have suppressed colour indices. Due to the antisymmetric colour index and the identity $\overline{\Psi}\chi^{c}=\overline{\chi}\Psi^{c}$ the couplings $\kappa_{1ijk}$ and $\kappa_{2ijk}$ are necessarily antisymmetric in down type quark flavour.

Other operators in addition to the ones of eq. (\ref{eq:eftlag}) are allowed by the symmetries of the theory e.g. $\overline{d_{Ri}}X_{1L}\overline{X_{2L}}d_{Rj}$. For simplicity we take such operators to be sufficiently suppressed compared to those of eq. (\ref{eq:eftlag}) as to be negligible. Even if present these would not qualitatively alter the outcomes below.  

The decays of the heavier $X_{\alpha}$ --- from now taken to be $X_{2}$ --- can be CP violating and if they occur out-of-equilibrium will lead to a baryon asymmetry. However, as we will see below, scatterings of the form $u_{i}X_{\alpha L} \to \overline{d_{j}}\overline{d_{k}}$ also violate CP: we will study the effects of this CP violation on the final asymmetry.

The above Lagrangian contains 28 physical phases if one considers couplings to all possible flavour combinations. As a demonstration and in order to simplify the analysis we will restrict ourselves to considering only couplings to the first generation up quark and second and third generation down quarks, $s$ and $b$. From now on we will suppress flavour indices on the above couplings. After rephasing we are left with two physical phases. For simplicity we set:
	\begin{equation}
	\kappa_{1}=e^{i\pi/2}|\kappa_{1}|,
	\end{equation} 
while we set the other phase (of $\kappa_{3}$) to zero. This choice for the phases leads to the largest CP violation: if the final asymmetry exceeds the observed one the asymmetry can always be reduced by decreasing the $\kappa_{1}$ phase. The other couplings are taken to be real.

\subsection{Decays}

The tree level decay rate $\Gamma(X_{\alpha} \to usb)+\Gamma(X_{\alpha} \to \overline{usb})$ is:
	\begin{equation}
	\Gamma_{\alpha A}=\frac{ |\kappa_{\alpha}|^{2}(M_{X\alpha})^{5} }{ 512\pi^{3} },
	\end{equation}
where we have ignored the masses of the final state particles. $X_{2}$ has an additional decay channel with rate $\Gamma(X_{2} \to X_{1}u\overline{u})$:
	\begin{equation}
	\Gamma_{2B}=\frac{ |\kappa_{3}|^{2}(M_{X2})^{5} }{ 1024\pi^{3} }
	\end{equation}
where we have ignored the masses of the final state particles (the full integral expression for massive $X_{1}$ can be found in appendix~\ref{sec:crosssections}). 

\begin{figure}[h]
\begin{center}
\includegraphics[width=300pt]{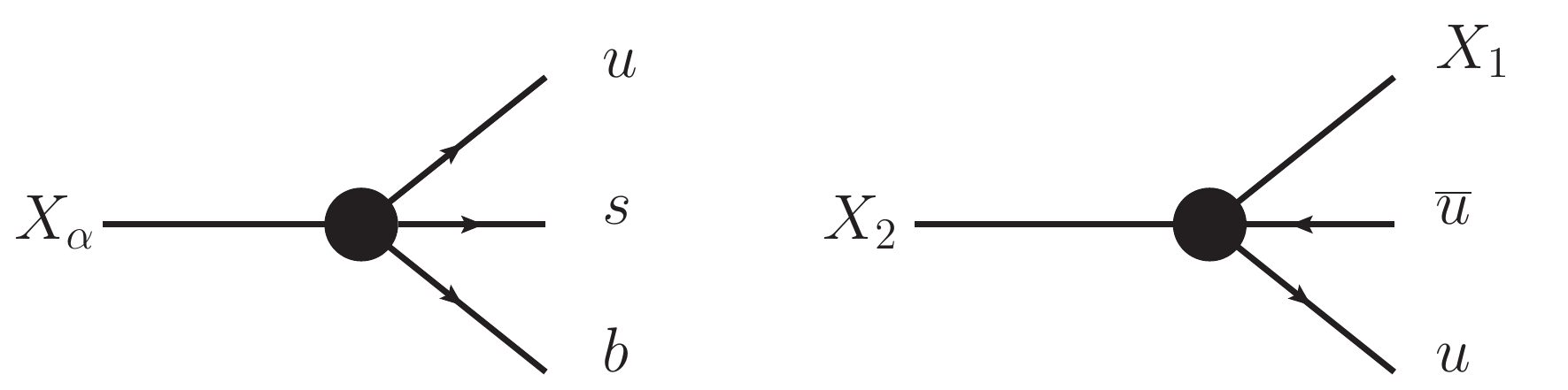}
\includegraphics[width=200pt]{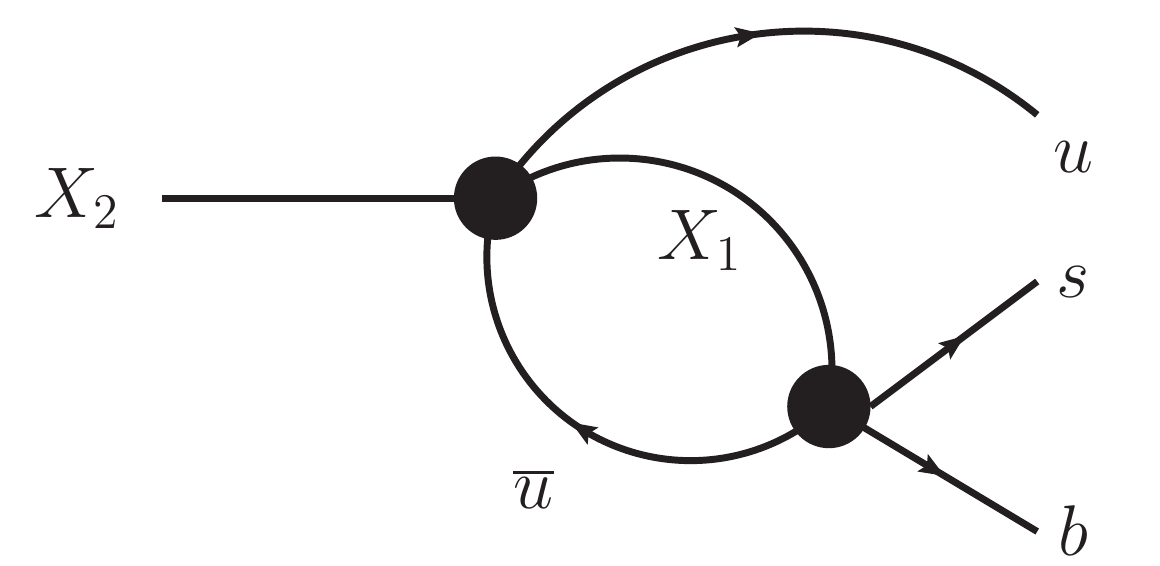}
\end{center}
\caption{Tree and loop level decays for $X_{1}$ and $X_{2}$. The particles in the loop for the lower diagram can go on-shell: interference between this and the tree diagram leads to CP violation in the $X_{2}$ decay. The $X_{2}$ in the loop in the analogous $X_{1}$ decay diagram cannot go on-shell and consequently there is no CP violation in the $X_{1}$ decay.}
\label{fig:decays}
\end{figure}

CP violation in the decays of $X_{2}$ arises from the interference between the tree and one loop diagrams depicted in figure~\ref{fig:decays}. We parameterise the CP violation in the $X_{2}$ decay in the following way:
	\begin{align}
	\Gamma(X_{2} \to usb) = \frac{1}{2}(1+\epsilon_{D})\Gamma_{2A}, \\
	\Gamma(X_{2} \to \overline{usb}) = \frac{1}{2}(1-\epsilon_{D})\Gamma_{2A}.
	\end{align}
The CP violation can be calculated in terms of the underlying parameters of the theory by using the Cutkosky rules to extract the imaginary component of the interference term~\cite{Peskin:1995ev}. The results are shown in figure~\ref{fig:cpdecay} and further details can be found in appendix~\ref{sec:crosssections}. 
\begin{figure}[h]
\begin{center}
\includegraphics[width=300pt]{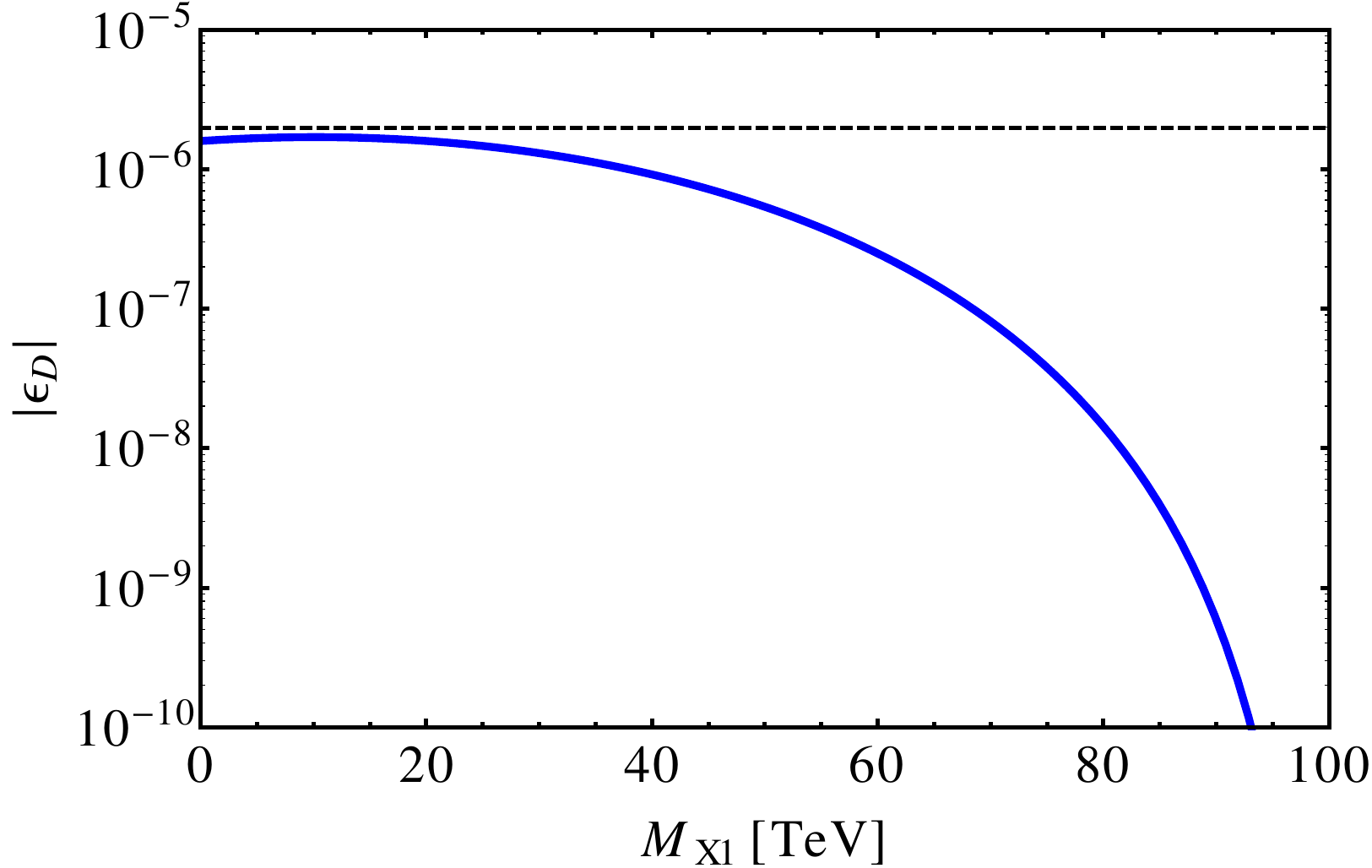}
\end{center}
\caption{CP violation in the decay of $X_{2}$ with $M_{X2}=100$ TeV and $\kappa_{a}=10^{-14}$ GeV$^{-2}$ as a function of $M_{X1}$. The horizontal dashed line corresponds to $\epsilon_{D}=\frac{\kappa_{a}}{16\pi}M_{X2}^{2}$.}
\label{fig:cpdecay}
\end{figure}
From dimensional arguments one expects,
	\begin{equation}
	\epsilon_{D} \sim \frac{1}{16\pi}\frac{\mathrm{Im}[\kappa_{1}^{\ast} \kappa_{2} \kappa_{3}^{\ast}]}{|\kappa_{2}|^{2}} M_{X2}^{2} \sim \frac{\kappa}{16\pi} M_{X2}^{2} ,
	\end{equation}
in the massless limit for $X_{1}$, where the second relation follows for couplings of a similar magnitude $\mathcal{O}(\kappa_{a}) \sim \kappa$ and an order one phase. This is confirmed by our detailed calculation, as can be seen in figure \ref{fig:cpdecay}.

One must also take into account CP violation in the scattering process $usb \to \overline{usb}$ mediated by an $X_{2}$ with the real intermediate part of the scattering subtracted~\cite{Kolb:1979qa}. We define this rate as:
	\begin{align}
	W(usb\to \overline{usb})=(1+\epsilon_{OS})W_{OS}
	\end{align}
where the CP conjugate can be found by taking $\epsilon_{OS} \to -\epsilon_{OS}$. The unitary condition (\ref{eq:eqcond}) --- taking $usb$ as the initial state and summing over all possible final states --- enforces the relation:
	\begin{equation}
	\epsilon_{OS}W_{OS}=\frac{1}{2}\epsilon_{D}n_{2}^{eq}\Gamma_{2A}.
	\end{equation}
This CP violating rate balances the CP violation in the decays when $X_{2}$ is in thermal equilibrium~\cite{Kolb:1979qa} and is included in our Boltzmann equations below. The CP symmetric washout rates $W(usb\to\overline{usb})$, mediated by off shell $X_{1}$ and $X_{2}$, are $\mathcal{O}(\kappa^{4})$ and are negligible for the parameters we are interested in below: these have been omitted from the Boltzmann equations.

\subsection{Scatterings}
\begin{figure}[t]
\begin{center}
\includegraphics[width=300pt]{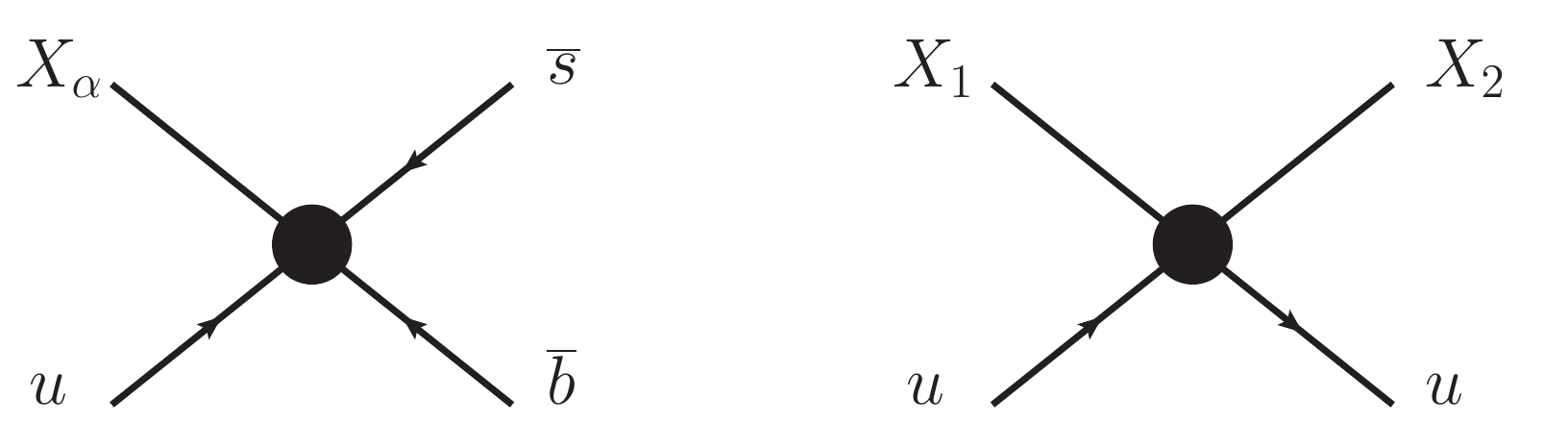}
\includegraphics[width=250pt]{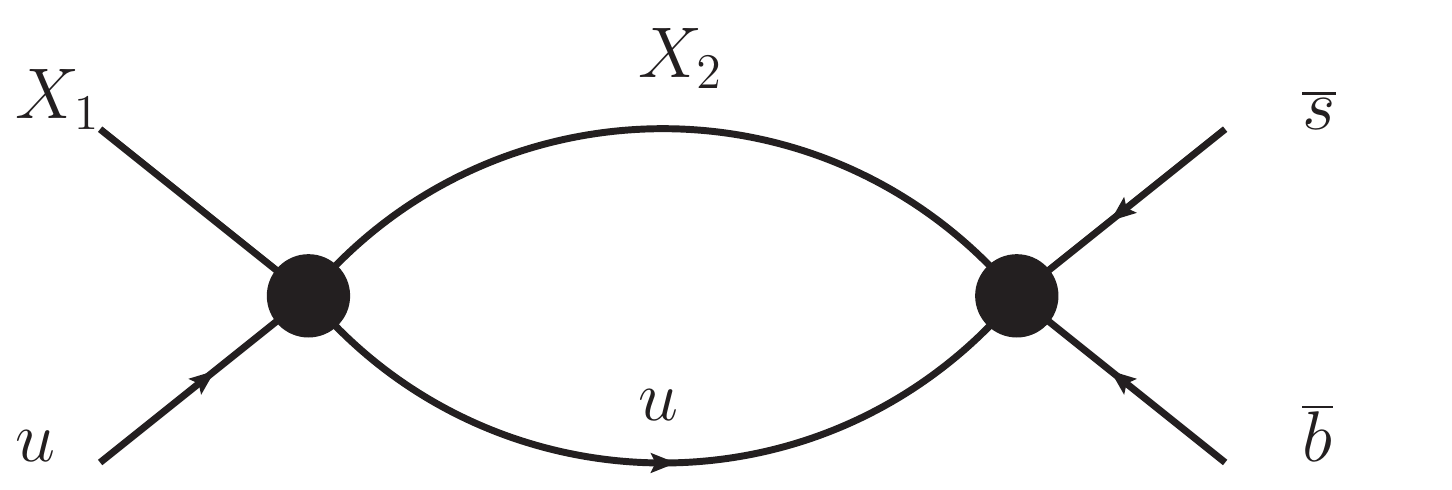}
\end{center}
\caption{Example tree and loop level scatterings leading to CP violation. Further loop diagrams exist with e.g. $X_{1}$ and $X_{2}$ interchanged.}
\label{fig:annihilation}
\end{figure}

Next we turn our attention to the scatterings. CP violation arises due to the interference between the tree and one-loop diagrams such as those depicted in figure~\ref{fig:annihilation}. We parameterise the relevant CP violating equilibrium collision terms as:
	\begin{align}
	W(u+X_{1} \to \overline{s}+\overline{b})=(1+\epsilon_{1})W_{1},  \\
	W(u+X_{2} \to \overline{s}+\overline{b})=(1+\epsilon_{2})W_{2},  \\
	W(u+X_{1} \to u+X_{2})=(1+\epsilon_{3})W_{3}, 
	\end{align}
where the CP conjugate interaction rate can be found by making the substitution $\epsilon_{i} \to -\epsilon_{i}$. These interactions will contribute to the baryon asymmetry. The unitarity conditions for these interactions give:\footnote{The unitarity condition for couplings to all possible quark flavour combinations is discussed in appendix~\ref{sec:unit}.}
	\begin{align}
	\epsilon_{1}W_{1}+\epsilon_{2}W_{2}=0, \\
	\epsilon_{1}W_{1}+\epsilon_{3}W_{3}=0,  \\
	\epsilon_{2}W_{2}-\epsilon_{3}W_{3}=0.  
	\end{align}
CP conserving (to the order we are working) collision terms are also present; we denote them in the following way:
	\begin{align}
	W(s+X_{1} \to \overline{u}+\overline{b})=W_{4},  \\
	W(s+X_{2} \to \overline{u}+\overline{b})=W_{5},  \\
	W(X_{1}+X_{1} \to u+\overline{u})=W_{6},	 \\
	W(X_{2}+X_{2} \to u+\overline{u})=W_{7},	 \\
	W(X_{1}+X_{2} \to u+\overline{u})=W_{8}. 
	\end{align}
We calculate the relevant cross sections and find the thermal averaged cross sections numerically by making use of the single integral formula \cite{Edsjo:1997bg}:
	\begin{align}
	\langle v \sigma(ij \to kl) \rangle = \frac{g_{i}g_{j}T}{8\pi^{4}n_{i}^{eq}n_{j}^{eq}} \int_{(m_{i}+m_{j})^{2}}^{\Lambda^{2}}p_{ij}E_{i}E_{j}v\sigma K_{1}\left(\frac{\sqrt{\hat{s}}}{T}\right) d\hat{s},
	\end{align}
where $\hat{s}$ is the centre-of-mass energy squared, $p_{ij}$ is the initial centre-of-mass momentum, $K_{1}(x)$ is the modified Bessel function of the second kind of order one and $\Lambda$ is the effective theory cut-off. The expressions for the cross sections can be found in appendix~\ref{sec:crosssections}.

\begin{figure}[t]
\begin{center}
\includegraphics[width=300pt]{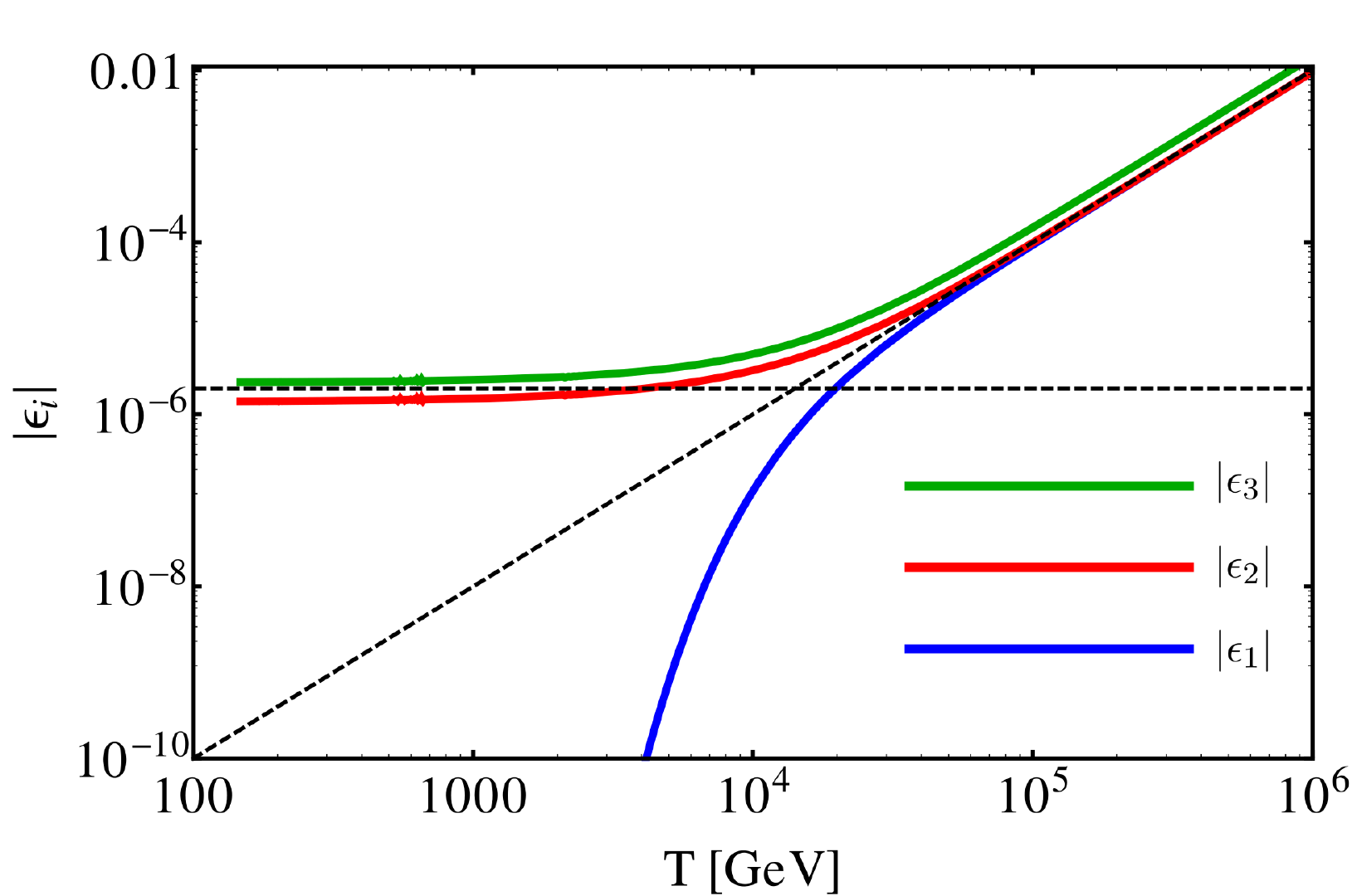}
\end{center}
\caption{CP violation in the scatterings with $M_{X2}=100$ TeV, $M_{X1}=50$ TeV, $\kappa_{a}=10^{-14}$ GeV$^{-2}$. The horizontal dashed line corresponds to $\epsilon=\frac{\kappa_{a}}{16\pi}M_{X2}^{2}$ and the diagonal dashed line corresponds to $\epsilon=\kappa_{a}T^{2}$. Note the kinematic suppression of $|\epsilon_{1}|$ at $T \lesssim M_{X2}$. \label{fig:cpannihilation}}
\end{figure} 

The CP violation in the scatterings at the cross section level i.e. the difference between the cross section and its CP conjugate, $(\sigma-\overline{\sigma})$, can be calculated in terms of the underlying parameters using the Cutkosky rules~\cite{Peskin:1995ev}. To find the CP violation in the equilibrium rates one must then take a thermal average:
	\begin{align}
	\langle v (\sigma(ij \to kl)-\sigma(\overline{ij} \to \overline{kl)}) \rangle = \frac{g_{i}g_{j}T}{8\pi^{4}n_{i}^{eq}n_{j}^{eq}} \int_{(m_{i}+m_{j})^{2}}^{\Lambda^{2}}p_{ij}E_{i}E_{j}v(\sigma-\overline{\sigma}) K_{1}\left(\frac{\sqrt{\hat{s}}}{T}\right) d\hat{s}.
	\end{align}
The results for such a calculation are depicted in figure~\ref{fig:cpannihilation}. At high temperature $T \gg M_{X2}$, one expects on dimensional grounds the CP symmetric thermally averaged cross section to scale as,
	\begin{equation}
	\langle \sigma v \rangle \sim \kappa^{2} T^{2},
	\end{equation}
where $\kappa$ is the general scale of the relevant couplings. Similary the CP violation is expected to scale as:
	\begin{equation}
	\epsilon \sim \kappa T^{2}.
	\end{equation}
Unlike the CP asymmetry in the decays, the CP asymmetries in the scatterings are temperature dependent. Note $\epsilon_{1}$ is necessarily suppressed on kinematic grounds at low temperatures ($W_{1}$ does not depend on the heaviest particle mass, $M_{X2}$, while the CP violation is sensitive to $M_{X2}$ from the on-shell particles in the loop). The other CP violating parameters take values close to
	\begin{equation}
	\epsilon \sim \frac{\kappa}{16\pi}M_{X2}^{2},
	\end{equation}
at temperatures $T \ll M_{X2}$, as can be seen in figure~\ref{fig:cpannihilation}.

\section{Boltzmann equations}
\label{sec:boltzmanneq}
\subsection{Differential equations}
The simplifying approximation of Maxwell-Boltzmann statistics allows one to factor out the chemical potential from the overall reaction rate density. Let us define:
	\begin{equation}
	r_{\Psi i}\equiv \frac{n_{\Psi i}}{n_{\Psi i}^{eq}}=\mathrm{Exp}\left(\frac{\mu_{\Psi}}{T}\right), \quad \quad \overline{r_{\Psi i}}\equiv \frac{n_{\overline{\Psi i}}}{n_{\Psi i}^{eq}}=\mathrm{Exp}\left(\frac{\mu_{\overline{\Psi}}}{T}\right).
	\end{equation}
The non-equilibrium reaction rate for the process $ij\to kl$ is then given by: $W^{\mathrm{neq}}(ij \to kl)=r_{i}r_{j}W(ij \to kl)$. The Boltzmann equations can then be conveniently expressed using this notation. The Boltzmann equation for $n_{X1}$ is given by:\footnote{The decay rates which appear here are thermally averaged (see appendix~\ref{sec:appdec}).}$^{,}$\footnote{We will see below that the three flavours of right chiral down type quarks have the same chemical potentials in the temperature ranges we are interested in. This allows us to set $r_{d}=r_{s}=r_{b}$ in the following equations.}
	\begin{align}	
	\frac{dn_{X1}}{dt} = & - 3Hn_{X1}  
	+\frac{1}{2}\Gamma^{\rm th}_{1A}n_{X1}^{eq}\Big[(\overline{r_{u}r_{d}r_{d}}+r_{u}r_{d}r_{d})-2r_{X1}\Big] + \Gamma^{\rm th}_{2B}n_{X2}^{eq}\Big[r_{X2}-r_{X1}\Big] \nonumber \\
	&\quad+ W_{1}\Big[r_{d}r_{d}+\overline{r_{d}r_{d}}-r_{X1}r_{u}-r_{X1}\overline{r_{u}}  \Big] + W_{3}\Big[r_{X2}r_{u}+r_{X2}\overline{r_{u}}  -r_{X1}r_{u}-r_{X1}\overline{r_{u}}\Big] \nonumber \\
	&\quad+ 2W_{4}\Big[r_{d}r_{u}+\overline{r_{d}r_{u}} -r_{X1}r_{d}-r_{X1}\overline{r_{d}}  \Big]	+ 2W_{6}\Big[1-r_{X1}r_{X1}  \Big] + W_{8}\Big[1-r_{X2}r_{X1}  \Big]	\nonumber \\
	&\quad+ \epsilon_{3}W_{3}\Big[\overline{r_{d}r_{d}}-r_{d}r_{d}+r_{X2}\overline{r_{u}}-r_{X2}r_{u}\Big].	\label{eq:bmannx1}
	\end{align}
The Boltzmann equation for $n_{X2}$ is given by:
	\begin{align}
	\frac{dn_{X2}}{dt} & =  - 3Hn_{X2}+  \frac{1}{2}\Gamma^{\rm th}_{2A}n_{X2}^{eq}\Big[((\overline{r_{u}r_{d}r_{d}}+r_{u}r_{d}r_{d})-2r_{X2}\Big]+ \Gamma^{\rm th}_{2B}n_{X2}^{eq}\Big[r_{X1}-r_{X2}\Big] \nonumber \\
	&\quad+ W_{2}\Big[r_{d}r_{d}+\overline{r_{d}r_{d}}-r_{X2}r_{u}-r_{X2}\overline{r_{u}}  \Big]
	+ W_{3}\Big[r_{X1}r_{u}+r_{X1}\overline{r_{u}}  -r_{X2}r_{u}-r_{X2}\overline{r_{u}}\Big] \nonumber	 \\
	&\quad+ 2W_{5}\Big[r_{d}r_{u}+\overline{r_{d}r_{u}} -r_{X2}r_{d}-r_{X2}\overline{r_{d}}  \Big]+ 2W_{7}\Big[1-r_{X2}r_{X2}  \Big] + W_{8}\Big[1-r_{X2}r_{X1}  \Big]	\nonumber \\
        &\quad+ \frac{1}{2}\epsilon_{D}\Gamma^{\rm th}_{2A}n_{X2}^{eq}\Big[(\overline{r_{u}r_{d}r_{d}}-r_{u}r_{d}r_{d})\Big]+ \epsilon_{3}W_{3}\Big[r_{d}r_{d}-\overline{r_{d}r_{d}}+r_{X1}r_{u}-r_{X1}\overline{r_{u}}\Big].	
	\end{align}
The Boltzmann equation for the baryon-minus-lepton number, $n_{B-L}$, is given by:
	\begin{align}
	\frac{dn_{B-L}}{dt} = & - 3Hn_{B-L}
	+\frac{1}{2}\Gamma^{\rm th}_{1A}n_{X1}^{eq}\Big[\overline{r_{u}r_{d}r_{d}}-r_{u}r_{d}r_{d}\Big]+\frac{1}{2}\Gamma^{\rm th}_{2A}n_{X2}^{eq}\Big[\overline{r_{u}r_{d}r_{d}}-r_{u}r_{d}r_{d}\Big] \nonumber \\
	&\quad+ W_{1}\Big[r_{X1}\overline{r_{u}}-r_{X1}r_{u}+\overline{r_{d}r_{d}}-r_{d}r_{d}  \Big]	
	+ W_{2}\Big[r_{X2}\overline{r_{u}}-r_{X2}r_{u}+\overline{r_{d}r_{d}}-r_{d}r_{d}  \Big]	\nonumber \\
	&\quad+ 2W_{4}\Big[r_{X1}\overline{r_{d}}-r_{X1}r_{d}+\overline{r_{d}r_{u}}-r_{d}r_{u}  \Big]	
	+ 2W_{5}\Big[(r_{X2}\overline{r_{d}}-r_{X2}r_{d}+\overline{r_{d}r_{u}}-r_{d}r_{u})  \Big]	 \nonumber \\
	&\quad+ \epsilon_{3}W_{3}\Big[r_{X1}\overline{r_{u}}+r_{X1}r_{u}-r_{X2}\overline{r_{u}}-r_{X2}r_{u}\Big] \nonumber \\	
	&\quad+ \frac{1}{2}\epsilon_{D}\Gamma^{\rm th}_{2A}n_{X2}^{eq}\Big[2r_{X2}-(\overline{r_{u}r_{d}r_{d}}+r_{u}r_{d}r_{d})\Big].	
	\end{align}
The terms proportional to $\epsilon_{3}$ and $\epsilon_{D}$ are the source terms coming from scatterings and decays respectively. These lead to the generation of a baryon asymmetry once there is a departure from thermal equilibrium. Note we have used the unitarity conditions to express the source terms in a form in which one sees they are explicitly zero when all $r_{\Psi i}=1$ (thermal equilibrium)~\cite{Baldes:2014gca}. The other terms are the washout terms: these drive the solutions back to the equilibrium values.

We use the standard change of variable to obtain the Boltzmann equations in a radiation dominated universe in terms of temperature using:
	\begin{equation}
	H = \left( \frac{8 \pi^{3} }{ 90 } \right)^{1/2} g_{\rm eff}^{1/2}  \frac{T^{2}}{M_{Pl}} = \frac{1}{2t},
	\end{equation}
where $g_{\rm eff}$ counts the effective radiation degrees of freedom and $M_{Pl}\approx 1.22 \times 10^{19}$ GeV is the Planck mass \cite{earlyuniverse}. The total entropy is conserved in the absence of first order phase transitions in a radiation dominated universe and hence it is useful to normalise number densities to the entropy density $Y_{\psi} \equiv n_{\psi}/s$. The entropy density is,
	\begin{equation}
 	s=\frac{2\pi^{2}}{45}h_{\rm eff}T^{3},
	\end{equation}
where $h_{\rm eff}$ counts the entropic degrees of freedom \cite{earlyuniverse}. The resulting differential equations are of the form,
	\begin{equation}
	\frac{dY_{\psi}}{dT} = - \left( \frac{\pi}{45} \right)^{1/2}\frac{M_{Pl}h_{\rm eff}}{s^{2}g_{\rm eff}^{1/2}}\left(1+\frac{T}{4 g_{\rm eff}}\frac{dg_{\rm eff}}{dT}\right)C(\psi)
	\end{equation}
where $C(\psi)$ is the collision term for $\psi$, e.g. for $\psi=X_{1}$ it corresponds to the right hand side of eq. (\ref{eq:bmannx1}) excluding the expansion term proportional to $H$.

\subsection{Chemical potentials}
The chemical potentials of the quarks and leptons depend on the baryon and lepton asymmetries. Here we are interested in determining the chemical potentials of the right chiral up and down type quarks. We consider two regimes: above and below the electroweak phase transition. We approximate the transition to occur instantaneously at a temperature $T_{EW}$ at which point we take the SU(2) sphalerons to have switched off and the SM quarks and leptons to have gained masses through the Higgs mechanism.
\subsubsection{Above $T_{EW}$}
In determining the chemical potentials in terms of $n_{B-L}$ in this regime one takes into account: (i) the SU(3) sphalerons, (ii) the SU(2) sphalerons, (iii) the SM Yukawa interactions, (iv) conservation of the weak hypercharge. The analysis is the same as for the usual leptogenesis~\cite{Buchmuller:2000as}. The chemical potential of the right chiral up type quark is given by:
	\begin{equation}
	\mu_{u R} = -\frac{10}{79}\frac{n_{B-L}}{T^{2}}.
	\end{equation}
The right chiral down type quark chemical potential is given by:
	\begin{equation}
	\mu_{d R} = \frac{38}{79}\frac{n_{B-L}}{T^{2}}.
	\end{equation}
The baryon asymmetry is related to $n_{B-L}$:
	\begin{equation}
	n_{B}=\frac{28}{79}n_{B-L}.
	\end{equation}
The chemical potentials of the up and down type quarks are therefore determined in terms of $n_{B-L}$ in this temperature regime. The final $n_{B}$ in this temperature regime is used as an initial condition for the Boltzmann equations at lower temperatures.
\subsubsection{Below $T_{EW}$} 
Below the electroweak phase transition the electroweak sphalerons have switched off and the weak hypercharge is no longer conserved. However, lepton number is now conserved and the overall electric charge must still vanish. The fermionic fields present are listed in table~\ref{tab:dof}. The baryon number asymmetry is approximately given by~\cite{Buchmuller:2000as}:
	\begin{equation}
	n_{B} =\frac{T^{2}}{6}B,
	\end{equation}
where $B=2N_{u}\mu_{u}+2N_{d}\mu_{d}$ and $N_{u}$ ($N_{d}$) is the number of relativistic up (down) type quark flavours. The lepton number asymmetry is approximately given by~\cite{Buchmuller:2000as}:
	\begin{equation}
	n_{L}=\frac{T^{2}}{6}L,
	\end{equation}
where $L=2N_{e}\mu_{e}+3\mu_{\nu}$ and $N_{e}$ is the number of relativistic charged lepton generations.
\begin{table}
\begin{center}
\begin{tabular}{| l | l | l | l | }
\hline
Field & Relativistic degrees of freedom & chemical potential \\ \hline
$u$ & $6 \times N_{u}$ & $\mu_{u}$ \\ \hline
$d$ & $6 \times N_{d}$ & $\mu_{d}$ \\ \hline
$e$ & $2 \times N_{e}$ & $\mu_{e}$ \\ \hline
$\nu$ & $1 \times N_{\nu}$ & $\mu_{\nu}$ \\ \hline
\end{tabular}
\caption{SM fermionic fields and relativistic degrees of freedom for temperatures below the electroweak phase transition. We set $N_{\nu}=3$ and use a simple step function in determining $N_{u}$, $N_{d}$ and $N_{e}$. \label{tab:dof}}
\end{center}
\end{table}
The chemical potentials are determined using the following constraints:
\begin{itemize}
\item Net electric charge vanishes. Once the temperature drops below the mass of a field its number density is very rapidly suppressed, if the field is in chemical equilibrium, as is the case for the SM fermions. We approximate this by only counting fields with mass below the temperature in contributing to the net charge of the plasma. This yields a constraint:
	\begin{equation}
	2N_{u}\mu_{u} - N_{d}\mu_{d} -N_{e} \mu_{e} = 0.
	\end{equation} 
\item  Rapid interactions involving $W$ boson exchange give the following relation:
	\begin{equation}
	\mu_{u}+\mu_{e}-\mu_{d}-\mu_{\nu}= 0.
	\end{equation}
\item $Y_{L}\equiv n_{L}/s$ is constant after $T_{EW}$. This gives the constraint:
	\begin{equation}
	\mu_{\nu}=\frac{L-2N_{e}\mu_{e}}{3},
	\end{equation} 
	where $L$ is computed using the lepton asymmetry $Y_{L}$ at $T_{EW}$.
	\end{itemize}
Combining these three constraints gives the up quark chemical potential:
	\begin{equation}
	\mu_{u} = \left\{L+B\left[\frac{1}{3}+\frac{1}{2N_{d}}+\frac{1}{2N_{e}}\right]\right\}\times \left[1+\frac{3N_{u}}{N_{e}}+\frac{N_{u}}{N_{d}}+2N_{u}\right]^{-1}.
	\end{equation}
The down type chemical potential can neatly be expressed as:
	\begin{equation}
	\mu_{d} = \frac{B-2N_{u}\mu_{u}}{2N_{d}}.
	\end{equation}
In this temperature regime the chemical potentials of the quarks can therefore be determined in terms of $n_{B}$ and the net lepton number $n_{L}$ at the sphaleron switch-off temperature. 

\subsection{Numerical solutions}
\label{sec:numerical}
We solve the Boltzmann equations numerically using Mathematica~\cite{mathematica}. This involves evolving the three coupled ordinary differential equations from a suitably high temperature to a temperature at which the $B$ violation is no longer effective. The initial temperature is chosen to be high enough so the $2 \leftrightarrow 2$ interaction rates are well above the expansion rate. The initial conditions correspond to equilibrium distributions for $X_{1}$ and $X_{2}$ and zero $n_{B-L}$.\footnote{Any initial $B-L$ would be erased by the on-shell interactions involving the new particles at the UV completion scale --- even below this scale the effective $2\leftrightarrow2$ interactions are rapid enough to make the analysis independent of initial conditions so long as the post-inflation reheating temperature of the universe is sufficiently high.}
At the electroweak phase transition the sphalerons switch off and the lepton number remains constant. We then proceed to track $n_{B}$ rather than $n_{B-L}$ using the value of $n_{B}$ obtained at $T_{EW}$ as an initial condition.

Finally a dilution factor is applied to the final $Y_{B}$ if the $X_{2}$ and $X_{1}$ particles are sufficiently long lived. This dilution factor comes about because the $X_{2}$ and $X_{1}$ can come to dominate the energy density of the early universe. The universe switches from a radiation to a matter dominated epoch until the $X_{2}$ and $X_{1}$ decay. The decays of these particles then produces a large amount of entropy --- slowing down the rate at which the universe is cooling --- and lead to a departure from the assumption of constant entropy in the early universe. The dilution factor can be approximated as \cite{PhysRevD.31.681}:
	\begin{equation}
	d_{S}=\mathrm{Max}\left[1.8 h_{\rm eff}^{1/4} \frac{Y_{\alpha}|_{T \rm{fo}}M_{X \alpha}}{(\Gamma_{\alpha} M_{Pl})^{1/2} }, \; 1\right],
	\end{equation}
where $Y_{\alpha}|_{T \rm{fo}}$ is the density to entropy ratio of the decaying particle at freeze out and $\Gamma_{\alpha}$ is its decay rate. Being more abundant at freezeout and having a longer lifetime $X_{1}$ tends to dominate the dilution factor.

\begin{figure}[t]
\begin{center}
\includegraphics[width=350pt]{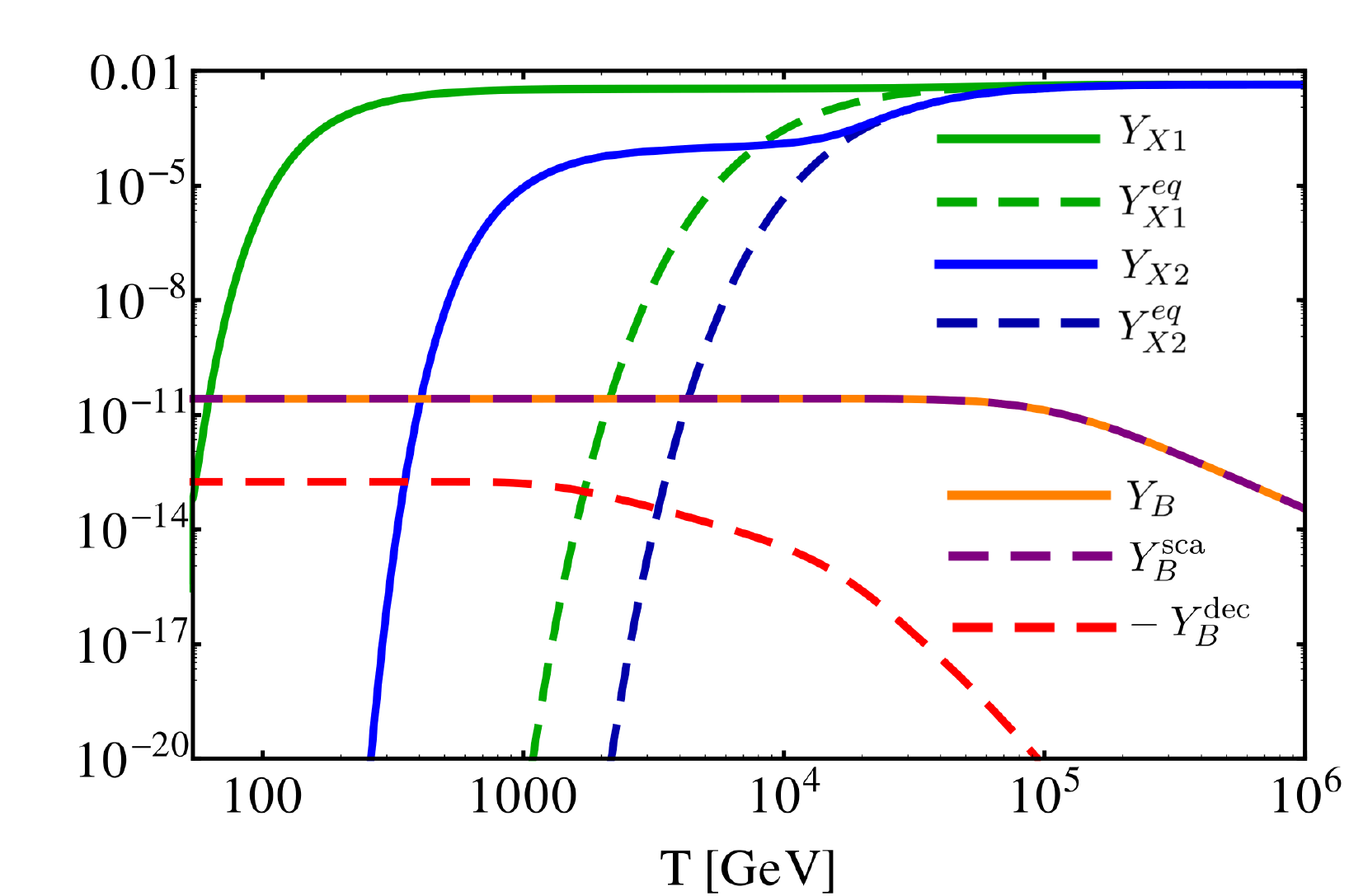}
\end{center}
\caption{Example solution to the Boltzmann equations with $M_{X2}=100$ TeV, $M_{X1}=50$ TeV, all $\kappa_{a}=10^{-16}$ GeV$^{-2}$ and number densities expressed in ratios to entropy $Y_{\Psi}\equiv n_{\Psi}/s$. The baryon asymmetry with CP violation only in scatterings (decays) $Y_{B}^{\mathrm{sca}}$ ($Y_{B}^{\mathrm{dec}}$) is also shown. CP violating scatterings are the dominant asymmetry generation mechanism in this example.}
\label{fig:exampleboltz}
\end{figure}

An example solution to the Boltzmann equations is shown in figure~\ref{fig:exampleboltz}. The cosmological history of the universe proceeds in the following way. At high temperature the $2 \leftrightarrow 2$ interactions are rapid, keeping the $n_{X1}$ and $n_{X2}$ close to their equilibrium values. Due to the expansion of the universe the particles are never exactly in equilibrium. As the temperature decreases the interaction rates drop and $r_{X1}$, $r_{X2}$ and $Y_{B}$ continue to increase. The size of the source term will depend not only on the CP violation but also on how far away the temperature is from $M_{X2}$ and $M_{X1}$: in the massless limit $r_{X1}=r_{X2}=1$ even in the absence of interactions (assuming a common $T$). Eventually, at $\Gamma_{\alpha} \sim H$, the decays take over. Excess $X_{2}$ and $X_{1}$ decay away and the $X_{2}$ decays also contribute to the baryon asymmetry. The freezeout temperature is determined by the coupling size; numerically we find the maximum asymmetry for freezeout at $T\sim M_{\alpha}$, i.e. before the number densities of $X_{1}$ and $X_{2}$ become Boltzmann suppressed.

\begin{figure}[t]
\begin{center}
\includegraphics[width=210pt]{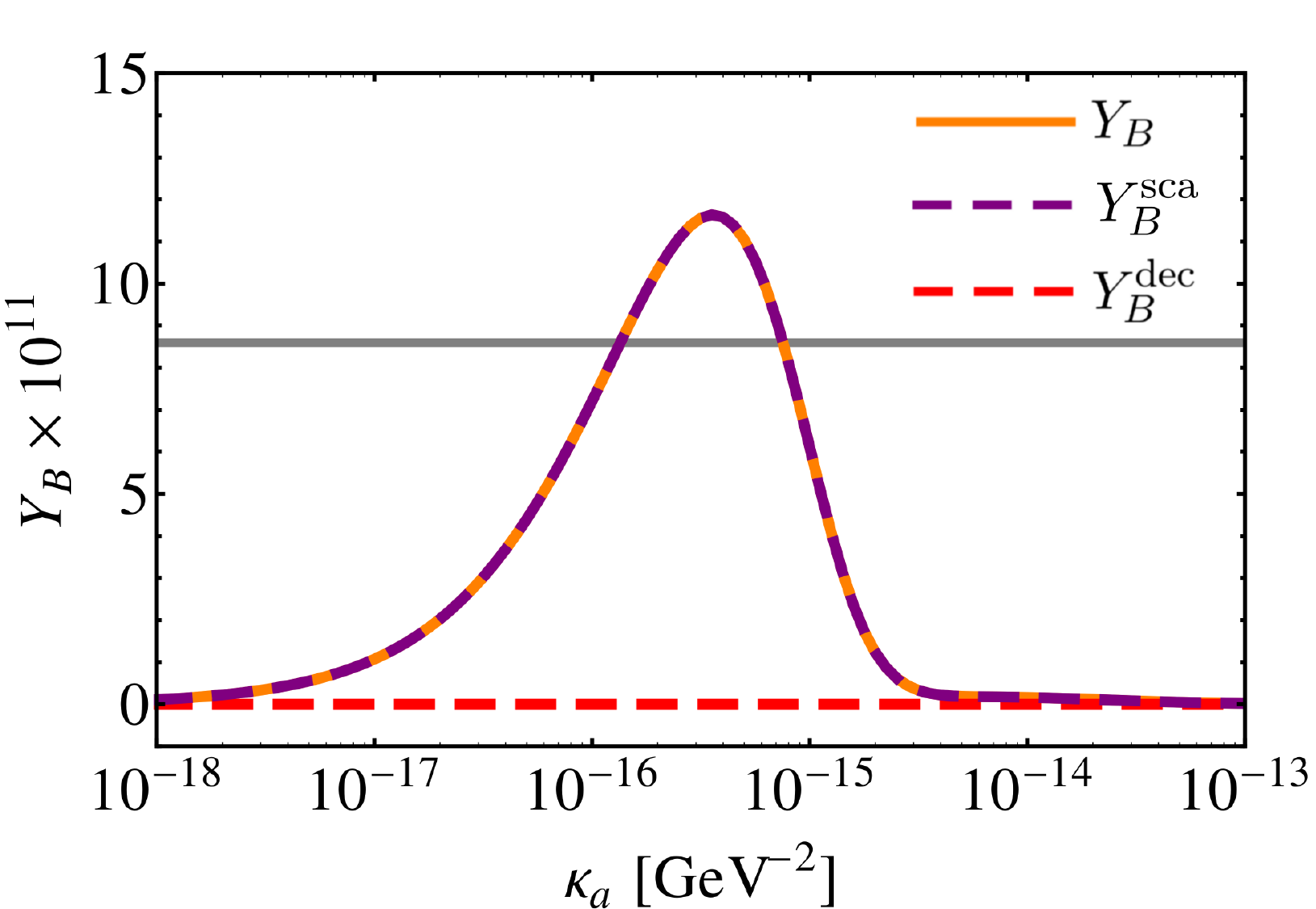}
\includegraphics[width=210pt]{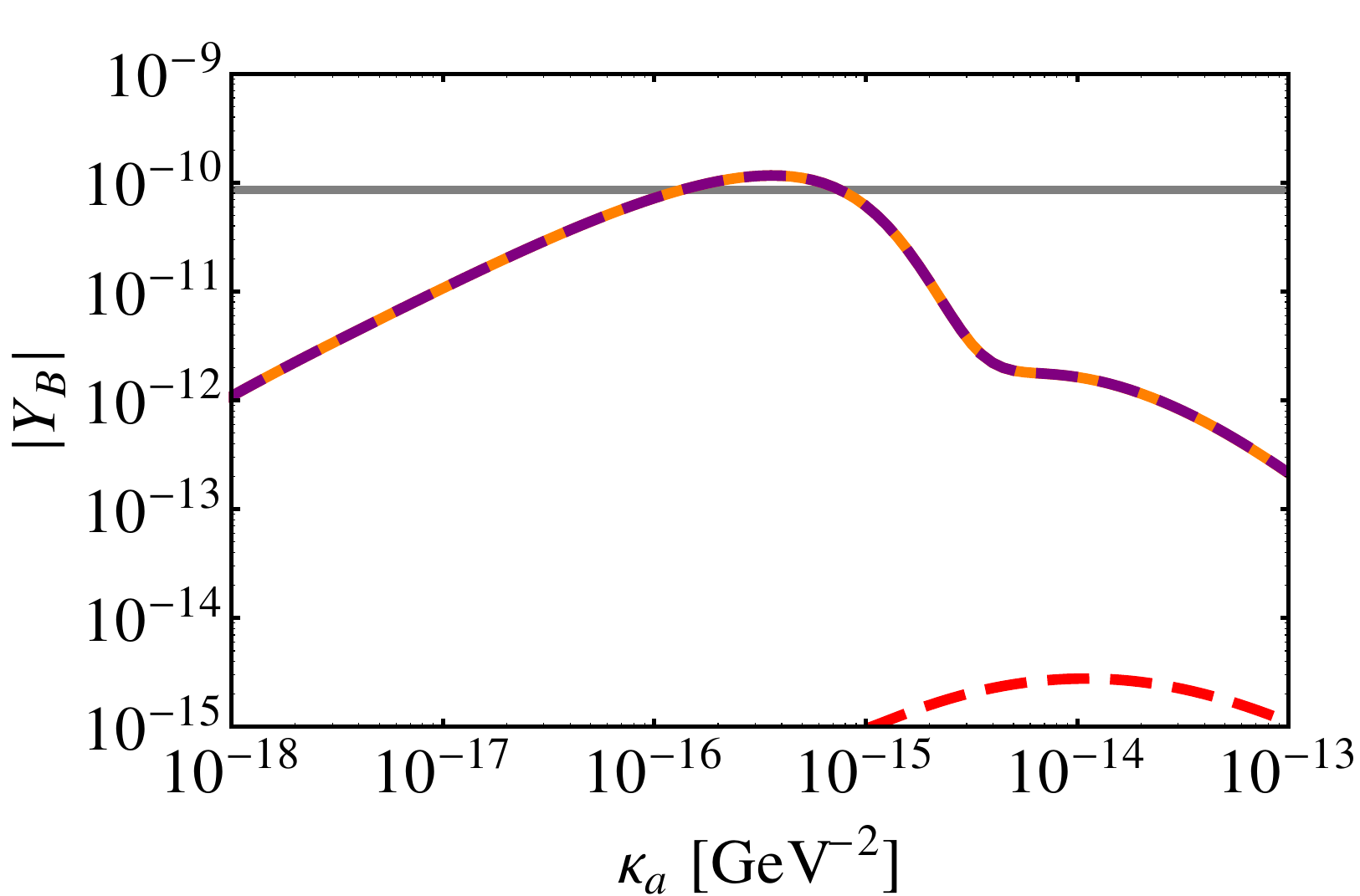}
\end{center}
\caption{\emph{Left}: the final baryon asymmetry as a function of the couplings (all set equal) $\kappa_{a}$, the masses have been set to $M_{X2}=100$ TeV and $M_{X1}=90$ TeV. The obseved value of the baryon asymmetry is indicated by a gray horizontal line and is reached for couplings in the range $10^{-16} \; \mathrm{GeV}^{-2} \lesssim \kappa_{a} \lesssim 10^{-15} \; \mathrm{GeV}^{-2}$. Also shown are the final asymmetries $Y_{B}^{\mathrm{sca}}$ ($Y_{B}^{\mathrm{dec}}$) calculated with CP violation only in the scatterings (decays). \emph{Right}: same as in the left plot but on a double logarithmic scale.}
\label{fig:couplingscan1}
\end{figure}

\begin{figure}[t]
\begin{center}
\includegraphics[width=210pt]{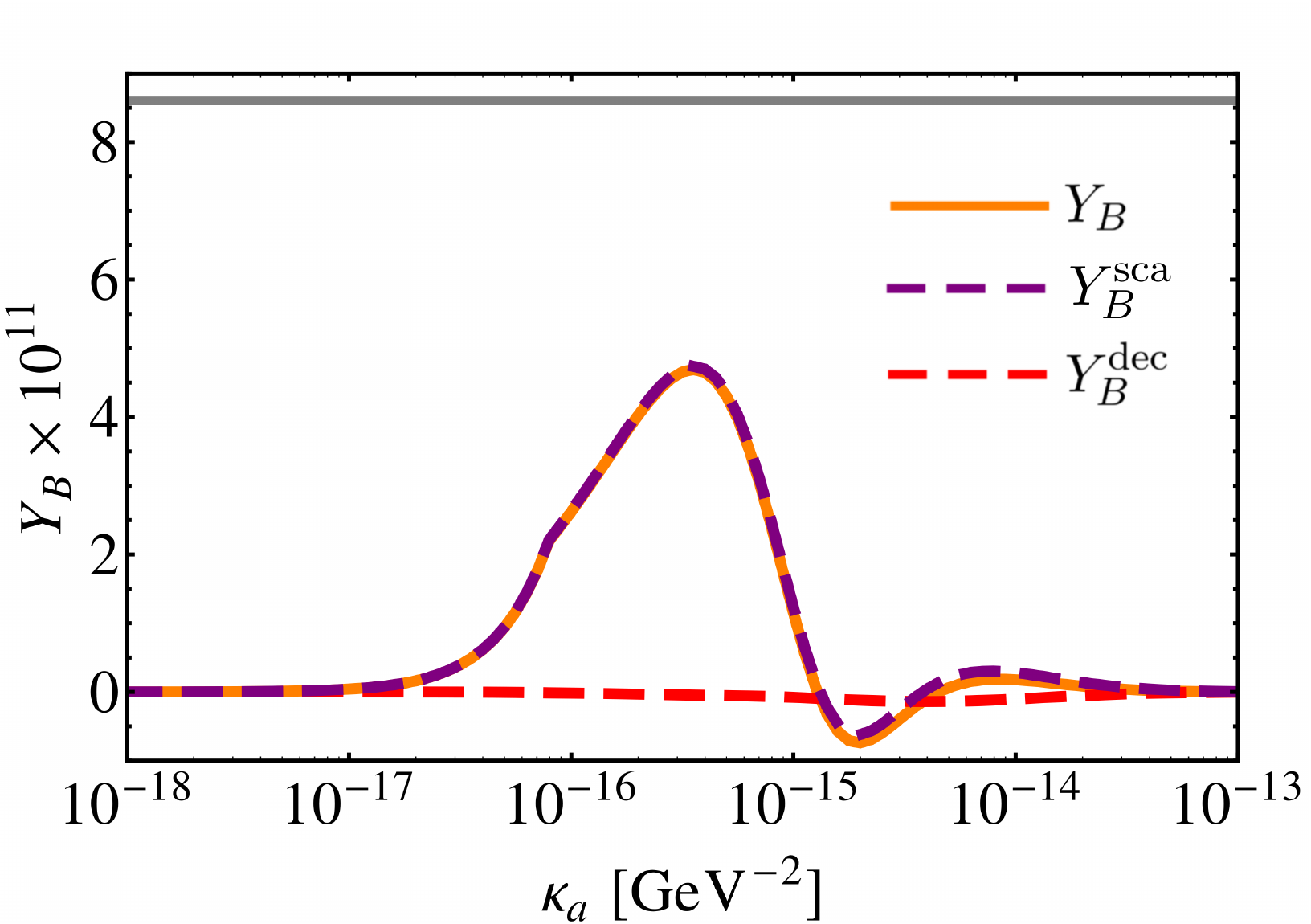}
\includegraphics[width=210pt]{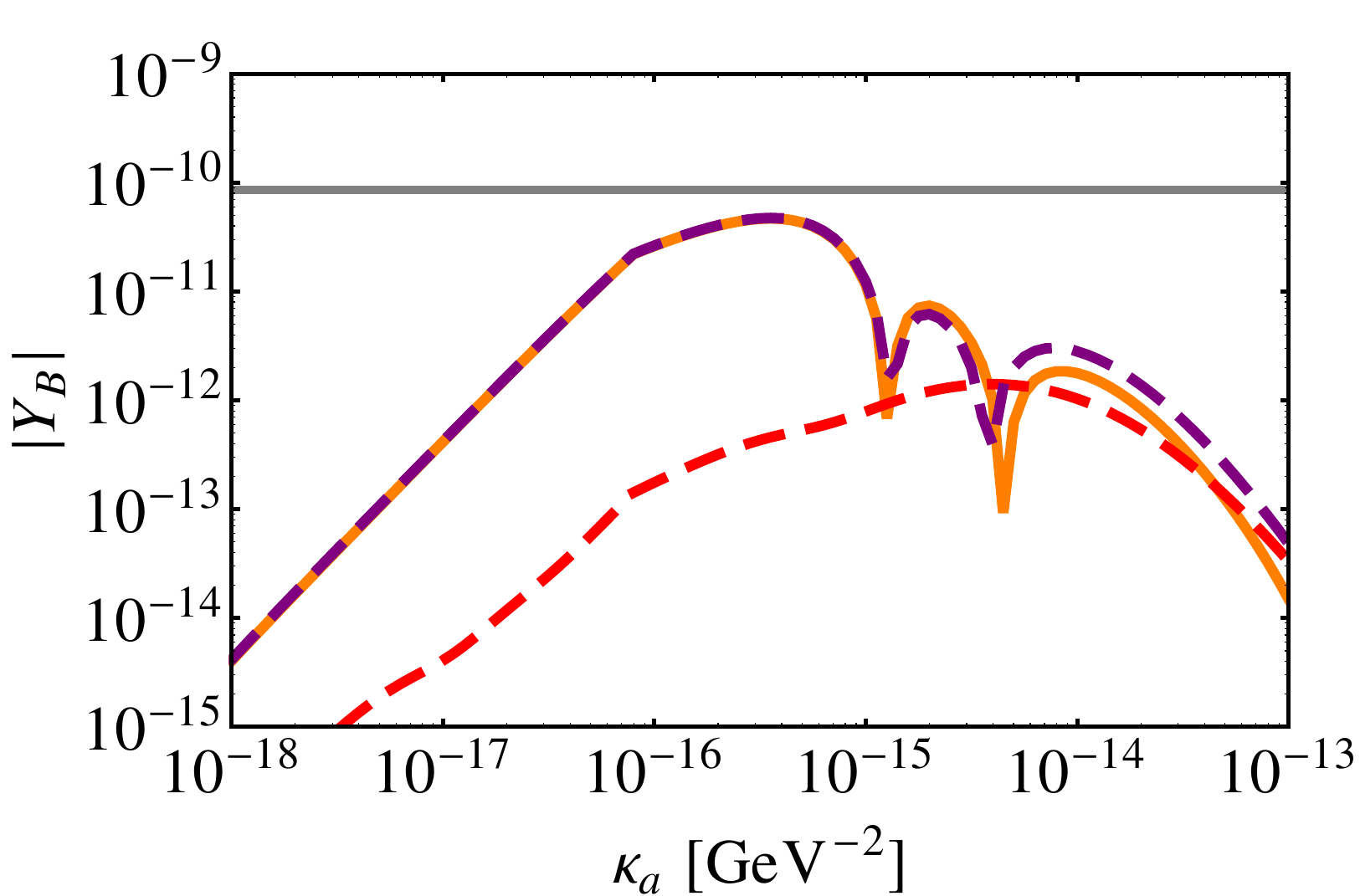}
\end{center}
\caption{Same as in figure~\ref{fig:couplingscan1} but with $M_{X1}=50$ TeV. The decays now play a more significant role but the final baryon asymmetry cannot match the observed value for any choice of couplings $\kappa_{a}$.}
\label{fig:couplingscan2}
\end{figure}

\begin{figure}[t]
\begin{center}
\includegraphics[width=210pt]{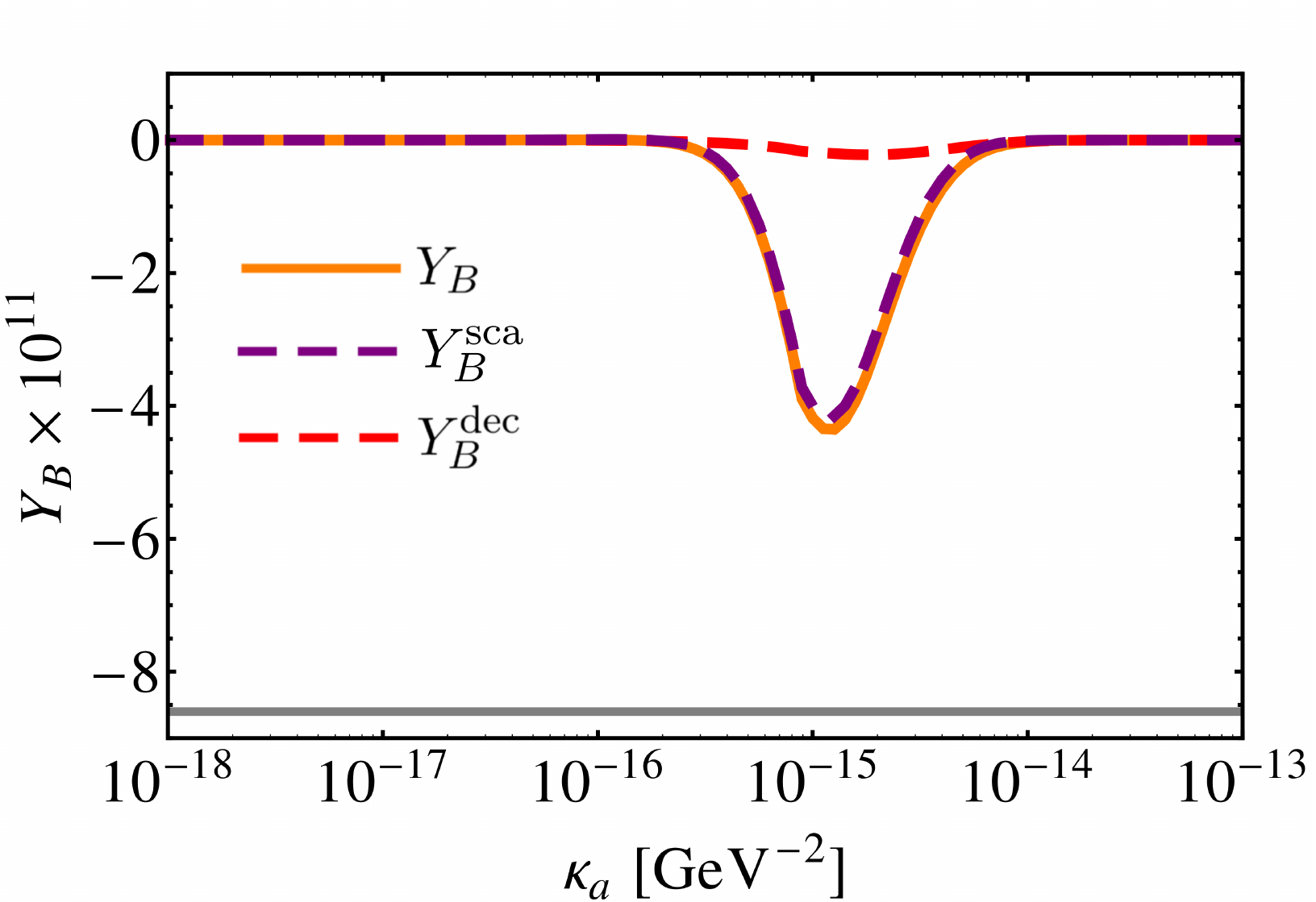}
\includegraphics[width=210pt]{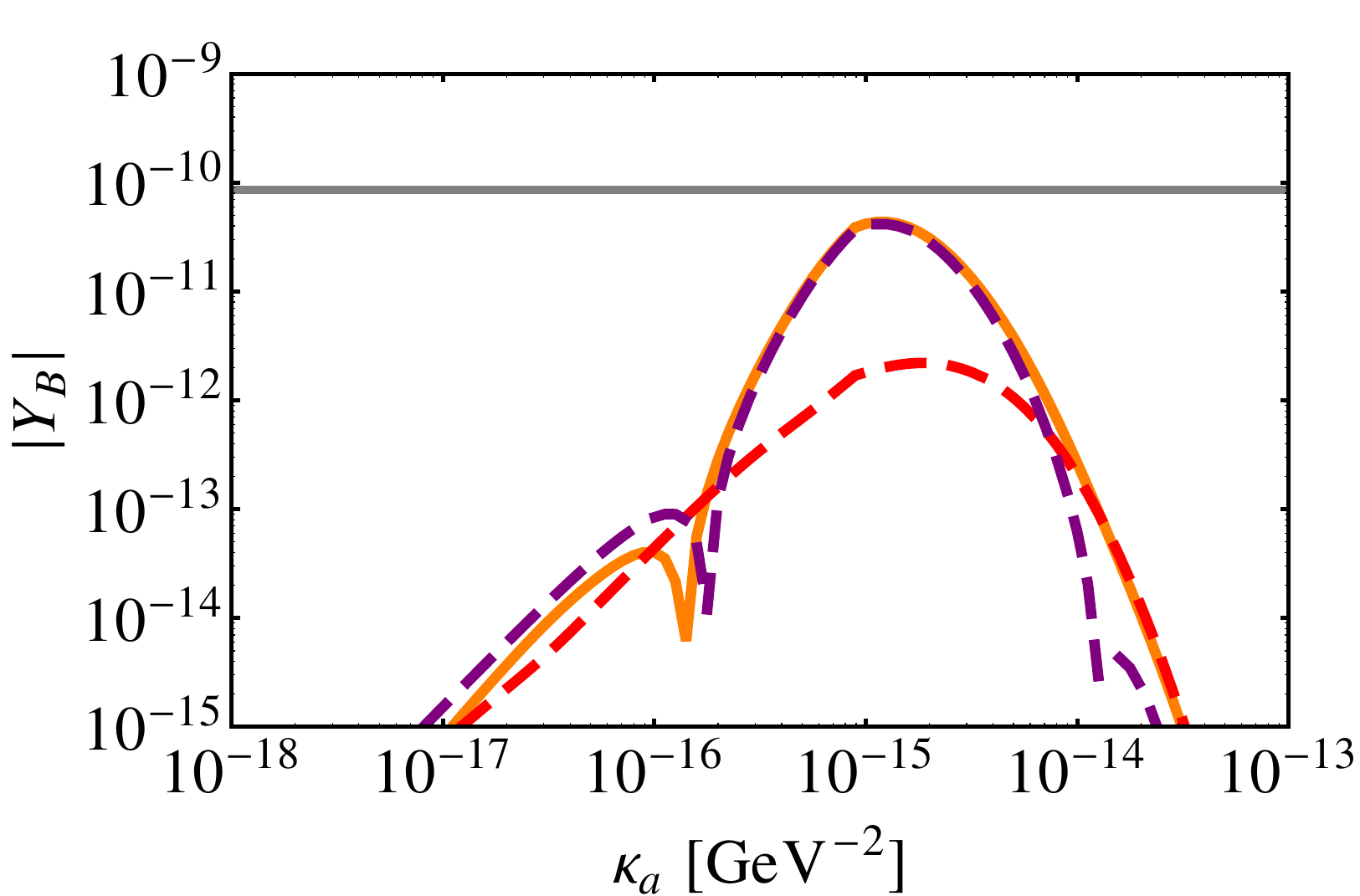}
\end{center}
\caption{Same as in figure~\ref{fig:couplingscan1} but with $M_{X1}=10$ TeV. Note the sign of the asymmetry can be changed by changing the sign of the CP violating phase. The horizontal gray line now indicates the magnitude of the observed asymmetry but with opposite sign. The CP violation in the decays is now close to maximal but the scatterings still dominate for $\kappa_{a} \lesssim 10^{-14}$ GeV$^{-2}$.}
\label{fig:couplingscan3}
\end{figure}

The CP violation in the decays and the scatterings increases $\sim \kappa$. However, as $\kappa$ increases the freeze out of the $2 \leftrightarrow 2$ interactions occurs closer and closer to the decay temperature, i.e. at $\Gamma_{\alpha} \sim H$, and the particles are kept closer to thermal equilibrium. The resulting effect on $Y_{B}$ is shown in figures~\ref{fig:couplingscan1} -- \ref{fig:couplingscan3} in which the final asymmetry first increases as the couplings increase but eventually becomes suppressed. 

Also shown in figures~\ref{fig:couplingscan1} -- \ref{fig:couplingscan3} is the domination of the CP violating scatterings in determing the final $Y_{B}$ for the majority of the parameter space. This is because the CP violation due to the scatterings --- which scales as $\kappa T^{2}$ --- can be relatively high at the freezeout temperature. This balances the relatively small departure from equilibrium just prior to freezeout, when the $2 \leftrightarrow 2$ interactions generate the majority of the asymmetry, compared with the departure from equilibrium at $\Gamma_{2} \sim H$, when the decays contribute to the asymmetry. The CP violation in the decays is typically much smaller than the CP violation in the scatterings at freeze out --- at least for couplings small enough for a significant departure from equilibrium to take place. As $M_{X1}$ is decreased the CP violation in decays becomes more important. The reasons are twofold: smaller $M_{X1}$ results in a greater $|\epsilon_{D}|$ and in larger washout effects after $X_{2}$ freezes out for kinematic reasons.

Note that the scaling of the CP violation with temperature is different in leptogenesis. The dimensionless couplings in leptogenesis mean the CP violation in scatterings will not grow as strongly with temperature, as it does here, but remain mostly constant~\cite{Pilaftsis:2003gt,Pilaftsis:2005rv,Nardi:2007jp,Davidson:2008bu,Fong:2010bh}. This explains why the CP violating scatterings can play a crucial role in the neutron portal but only a negligible role in resonant leptogenesis in determining the final $Y_{B}$~\cite{Pilaftsis:2003gt}.

The precision of these calculations can be improved by taking into account departures from kinetic equilibrium, quantum statistics and thermal masses~\cite{Hannestad:1999fj,Giudice:2003jh,Basboll:2006yx,Garayoa:2009my,HahnWoernle:2009qn}. The corrections are expected to be at most $\mathcal{O}(1)$, as in the case for leptogenesis. We leave the inclusion of such effects to further work.

\section{Constraints}
\label{sec:constraints}
Constraints on the neutron portal have previously been discussed in ref.~\cite{Cheung:2013hza}. The decay of relic particles after $t \sim 1$ s can disrupt big bang nucleosynthesis (BBN)~\cite{Kawasaki:2004yh,Kawasaki:2004qu,Jedamzik:2006xz}. Considering the $X_{1}$ lifetime one finds a constraint~\cite{Cheung:2013hza}:
	\begin{equation}
	\kappa_{1} \gtrsim \left( \frac{1 \; \text{TeV}}{M_{X1}} \right)^{5/2}10^{-18} \; \text{GeV}^{-2}. 
	\end{equation}
Similarly from the $X_{2}$ lifetime one requires~\cite{Cheung:2013hza}:
	\begin{equation}
	\kappa_{2} \; \text{or} \; \kappa_{3} \gtrsim \left( \frac{1 \; \text{TeV}}{M_{X2}} \right)^{5/2}10^{-18} \; \text{GeV}^{-2}, 
	\end{equation}
where we have ignored the final state $M_{X1}$ mass. The $\kappa_{3}$ bound becomes more stringent as $M_{X1}$ is increased.

\begin{figure}[t]
\begin{center}
\includegraphics[width=200pt]{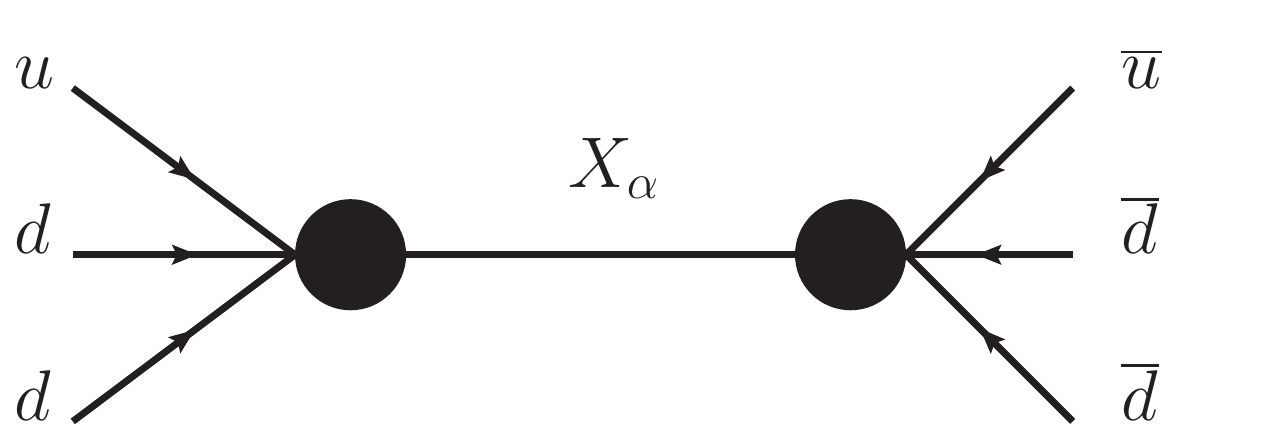}
\end{center}
\caption{Neutron-antineutron oscillation induced by operators of the form $\kappa \overline{X_{\alpha L}} d_{R} \overline{u_{R}^{c}} d_{R}$ or $\kappa \overline{X_{\alpha L}} d_{R} \overline{Q_{L}^{c}} Q_{L}$.}
\label{fig:nnbar}
\end{figure}

Operators of the form $\kappa \overline{X_{\alpha L}} d_{R} \overline{u_{R}^{c}} d_{R}$ or $\kappa \overline{X_{\alpha L}} d_{R} \overline{Q_{L}^{c}} Q_{L}$, where $X_{\alpha}$ is Majorana, are constrained by limits on neutron-antineutron oscillations (see figure~\ref{fig:nnbar}). The oscillation period is estimated as~\cite{Cheung:2013hza},
	\begin{equation}
\tau_{n-\overline{n}} \sim 3 \times 10^{8} \; \mathrm{s} \times\left( \frac{ M_{X\alpha} }{ 1 \; \mathrm{TeV} } \right) \left( \frac{  10^{-13} \; \mathrm{GeV}^{-2} }{ \kappa } \right)^{2} \left( \frac{  250 \; \mathrm{MeV} }{ \Lambda_{QCD} } \right)^{6}, 
\end{equation}
where we have approximated the nuclear matrix element with $\Lambda_{QCD}$. Comparison to the experimental limit from bound neutrons $\tau_{n-\overline{n}} \geq 2.4 \times 10^{8}$ s \cite{Abe:2011ky} (or from free neutrons $\tau_{n-\overline{n}} \geq 8.6 \times 10^{7}$ s~\cite{grenoble} --- which has smaller theoretical uncertainty) shows broad compatibility in the parameter range of interest. The operators we considered, however, couple only off-diagonally in down quark flavour. The oscillation period is therefore further suppressed and certainly does not pose any problems for the parameter choices we have been interested in above. Similar conclusions hold for the loop induced meson mixing operators \cite{Cheung:2013hza}. 

\begin{figure}[t]
	\begin{center}
	\includegraphics[width=150pt]{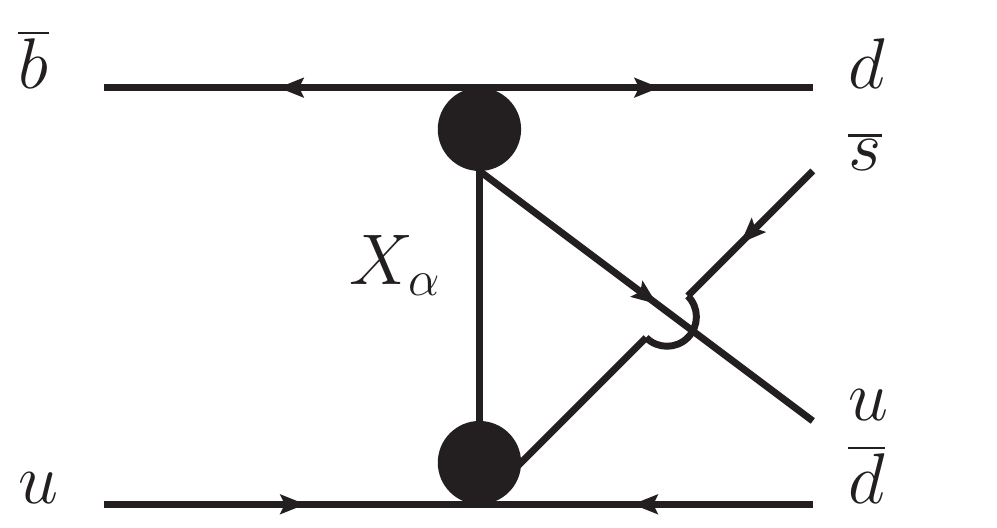}
	\end{center}
	\label{fig:meson}
	\caption{Meson decay $B^{+}\to\pi^{+}K^{0}$ due to the operators $\kappa \overline{X_{\alpha L}}u_{R}\overline{s_{R}^{c}}b_{R} + \kappa \overline{X_{\alpha L}}u_{R}\overline{s_{R}^{c}}d_{R}$.}
	\end{figure}

Flavour off-diagonal operators such as $\kappa \overline{X_{\alpha L}}u_{R}\overline{s_{R}^{c}}b_{R} + \kappa \overline{X_{\alpha L}}u_{R}\overline{s_{R}^{c}}d_{R}$ will lead to meson decays such as $B^{+}\to\pi^{+}K^{0}$ (see figure~\ref{fig:meson}). The contribution to the branching ratio is estimated as:
\begin{equation}
Br_{X}(B^{+}\to\pi^{+}K^{0}) \approx 10^{-38} \times\left( \frac{ 1 \; \mathrm{TeV} }{ M_{X\alpha} } \right)^{4} \left( \frac{ \kappa }{  10^{-12} \; \mathrm{GeV}^{-2} } \right)^{4}
\end{equation}
which is far below the experimental observation $Br(B^{+}\to\pi^{+}K^{0})=(2.3 \pm 0.07)\times 10^{-5}$ \cite{Beringer:1900zz,Aubert:2006gm,Duh:2012ie}. The most important constraints on this scenario therefore come from BBN and neutron-antineutron oscillations.

\section{UV completions}
\label{sec:uvcomplete}
One or more complex scalars transforming as $S_{\rho} \sim (3,1,4/3)$ under the SM gauge group may be added in order to UV complete this model. The high energy interaction Lagrangian takes the form: 
\begin{equation}
\Delta \mathcal L =  y_{\rho \alpha i} S_{\rho}^{\ast} \overline{X_{\alpha L}} u_{Ri}+ \lambda_{\rho ij}  S_{\rho} \overline{d_{Ri}^c}d_{Rj} + H.c,
\label{eq:lagrangian1} 
\end{equation}
where $y_{\rho \alpha i}$ and $\lambda_{\rho ij}$ are dimensionless couplings and the latter is antisymmetric in quark flavour. For couplings $y_{\rho \alpha i} \sim \lambda_{\rho ij} \sim \mathcal{O}(1)$ the effective operators of eq. (\ref{eq:eftlag}) are generated with $\kappa \sim 1/M_{S\rho}^{2}$, where $M_{S\rho}$ is the mass of the scalar $S_{\rho}$, once the $S_{\rho}$ are integrated out.

If the couplings are only to the $u$, $s$ and $b$ quarks two copies of $S_{\rho}$ must be added in order to obtain complex phases. However, in a realistic model --- allowing couplings to all possible quarks --- one finds two complex phases with only one $S_{\rho}$. The corresponding 28 complex phases of the low energy effective field theory are therefore not independent. This can lead to cancellations of phases in some of the interference terms and may reduce the overall CP violation. We leave investigation of such effects to further work.    

If we do not impose a global lepton number symmetry the $X_{\alpha}$ will mix with the SM neutrino degrees-of-freedom through Yukawa couplings of the form $\epsilon_{ab} \overline{l_{La}}\Phi_{b}^{\ast}X_{\alpha L}^{c}$, where $\epsilon_{ab}$ is the Levi-Civita symbol, $\Phi$ is the SM Higgs doublet and $l_{L}$ the SM lepton doublet. If the $X_{\alpha}$ are to be responsible for the observed neutrino masses, the mixing between the mostly-active light neutrinos and mostly-sterile heavy states must be of the order $\theta \sim \sqrt{m_{\nu}/M_{X\alpha}}$, where $m_{\nu}$ is the light neutrino mass. Consequently $S_{\rho}$ mediated proton decay $p\rightarrow K^+ \nu$ will occur with a partial lifetime estimated as:
\begin{align}
\tau_{p\rightarrow K^+ \nu} &\sim  \frac{10^{13}  \; \text{years}}{|y_{\rho \alpha u}\lambda_{\rho ds}|^{2}}\left( \frac{250 \; \text{MeV}}{\Lambda_{QCD}} \right)^{5} \left( \frac{0.05 \; \text{eV}}{m_{\nu}} \right) \left( \frac{M_{X\alpha}}{10 \; \text{TeV}} \right) \left( \frac{M_{S\rho}}{10^{7} \; \text{GeV}} \right)^{4}. 
\end{align}
The partial proton lifetime will conflict with the experimental limit $\tau_{p\rightarrow K^+ \nu} \gtrsim 2.3\times10^{33}$ years~\cite{Kobayashi:2005pe} for the parameter range of interest if the $X_{\alpha}$ are responsible for the neutrino mass through the usual seesaw mechanism.

Furthermore Yukawa couplings of the form $\epsilon_{ab} \overline{l_{La}}\Phi_{b}^{\ast}X_{\alpha L}^{c}$ would lead to leptogenesis style interactions and subsequent complications in the analysis of the cosmological history. The non-zero neutrino masses can be explained, even with the global lepton number symmetry we have imposed, by introducing right chiral singlets $N_{R}$ with Yukawa couplings $\epsilon_{ab} \overline{l_{La}}\Phi_{b}^{\ast}N_{R}$: these give the SM neutrinos Dirac masses once the SM Higgs gains a vacuum expectation value. Majorana masses for the $N_{R}$ are, of course, forbidden.

In this section we have briefly discussed UV completions for the the neutron portal effective operators of section \ref{sec:model}. Other neutron portal type operators are possible and we note that similar conclusions for their UV completions, in particular regarding proton decay if no global lepton number conservation is imposed, also hold.

\section{Conclusion}
We have studied the effects of CP violation in $2 \leftrightarrow 2$ interactions for baryogenesis. As a case study we have taken a neutron portal model in which two Majorana fermions $X_{1}$ and $X_{2}$ are coupled to the neutron operator. In order to find the final baryon asymmetry, we calculated all the relevant decay rates and $2 \leftrightarrow 2$ scattering rates and solved the Boltzmann evolution equations numerically. From dimensional grounds the CP violation in the decays scales as $\kappa M_{X2}^{2}$ where $\kappa$ is the order-of-magnitude of the relevant couplings. The CP violation in the scatterings scales as $\kappa M_{X2}^{2}$ at low temperatures but grows as $\kappa T^{2}$ for $T \gg M_{X}$. Consequently CP violating $2 \leftrightarrow 2$ scatterings can play a crucial role at high temperature even when the departure-from-equilibrium is relatively small. Indeed for many areas of the parameter space the CP violating scatterings play a dominant role in determining the final baryon asymmetry. This is to be contrasted with leptogenesis --- in which the CP violating scatterings do not scale as strongly with temperature --- and have only a negligible effect on the final asymmetry~~\cite{Pilaftsis:2003gt,Pilaftsis:2005rv,Nardi:2007jp,Davidson:2008bu,Fong:2010bh}.  We also discussed constraints on the model from experiments and BBN and discussed possible UV completions. The techniques learned can be applied to other baryogenesis scenarios in order to take into account the --- possibly dominant --- corrections due to CP violating scatterings. 

\acknowledgments{IB and AM were supported by the Commonwealth of Australia. NFB and RRV were supported in part by the Australian Research Council. KP was supported by the Netherlands Foundation for Fundamental Research of Matter (FOM) and the Netherlands Organisation for Scientific Research (NWO). Feynman diagrams drawn using Jaxodraw \cite{Binosi:2003yf}.}

\appendix

\section{Unitarity constraint for multiple quark generations}
\label{sec:unit}
Here we discuss the unitarity constraint in the case of couplings to all possible quark flavours and show that at least two $X_{\alpha}$ are required to obtain a non-zero source term from the CP violating scatterings.\footnote{For clarity, we only consider initial states $X_{\alpha}+d_{i}$ here; the arguments can easily be generalised to include $X_{\alpha}+u_{i}$ initial states.} We denote the equilibrium reaction rates as:
	\begin{align}
	W(X_{\alpha}+d_{i} \to \overline{u_{j}}+\overline{d_{k}})=(1+\epsilon_{\alpha i j k})W_{\alpha i j k}, \\
	W(X_{\alpha}+d_{i} \to X_{\beta}+d_{j})=(1+\zeta_{\alpha i \beta j})Z_{\alpha i \beta j}, \\
	W(u_{i}+d_{j} \to u_{k}+d_{l})=(1+a_{ijkl})A_{ijkl},
	\end{align}
where the CP conjugate can be found by taking $\epsilon_{\alpha i j k} \to -\epsilon_{\alpha i j k}$, $\zeta_{\alpha i \beta j} \to -\zeta_{\alpha i \beta j}$ or $a_{ijkl} \to - a_{ijkl}$. Note we have also included quark flavour changing interactions in order to make the argument more general. The unitarity constraint then yields:
	\begin{align}
	\sum_{\beta j}\zeta_{\alpha i \beta j}Z_{\alpha i \beta j}+\sum_{jk}\epsilon_{\alpha ijk}W_{\alpha ijk} = 0, \label{eq:unit1}\\
	\sum_{ml}a_{jkml}A_{jkml}+\sum_{\alpha i}\epsilon_{\alpha ijk}W_{\alpha ijk} = 0. \label{eq:unit2}
	\end{align} 
Now consider the case of only one Majorana $X_{\alpha}=X_{1}$. (The argument may be easily modified for Dirac $X_{1}$ and the conclusions remain unchanged.) Taking $r_{d}=r_{s}=r_{b}$ and $r_{u}=r_{c}=r_{t}$ the Boltzmann equation for the baryon asymmetry (excluding the decay terms) is:
	\begin{align}
	\frac{dn_{B}}{dt} + 3Hn_{B} = \sum_{ijk}W_{1ijk}\Big[(1-\epsilon_{1ijk})(\overline{r_{d}}\overline{r_{u}}+r_{X1}\overline{r_{d}})-(1+\epsilon_{1ijk})(r_{d}r_{u}+r_{X1}r_{d})\Big].
	\end{align}
Now consider only the source term:
	\begin{align}
	&-\sum_{ijk}\epsilon_{1ijk}W_{1ijk}\Big[\overline{r_{d}}\overline{r_{u}}+r_{X1}\overline{r_{d}}+r_{d}r_{u}+r_{X1}r_{d}\Big].
	\end{align}
A departure from equilibrium due to the expansion of the universe gives $r_{X1} \neq 1$ and $r_{u}=r_{d}=\overline{r_{u}}=\overline{r_{d}}=1$ and the source term may be written:
	\begin{equation}
	-(2+2r_{X})\sum_{i}\sum_{jk}\epsilon_{1ijk}W_{1ijk}=(2+2r_{X})\sum_{i}\sum_{j}\zeta_{1i1j}Z_{1i1j}=0,
	\end{equation}
where the first equality follows from eq.~(\ref{eq:unit1}) and the second as $\zeta_{\alpha i \alpha j}=-\zeta_{\alpha j \alpha i}$. Hence the generation of an asymmetry from scatterings with only a single $X_{\alpha}$ is not possible.

\section{Cross sections, decay rates and CP violation}
\label{sec:crosssections}
\tocless\subsection{Cross sections}
Here we collect the results of our calculations for the cross sections used in the above analysis. The sums over initial and final colours have been performed. The cross sections have been calculated in the centre-of-mass frame. $E_{a}$ and $E_{b}$ ($E_{c}$ and $E_{d}$) denote initial (final) state energies of the particles in the order listed. The initial momentum is denoted $p_{i}$ and the final momentum $p_{f}$. The centre-of-mass energy is $\sqrt{\hat{s}}$.
\begin{flalign}
&u+X_{\alpha} \to \overline{s}+\overline{b} & \nonumber \\
&E_{a}E_{b}\sigma v = \frac{3 |\kappa_{\alpha}|^{2} p_{f} }{2 \pi\sqrt{\hat{s}} } \left[ E_{a}E_{b}+p_{i}^{2}\right] \left[E_{c}E_{d}+p_{f}^{2}\right] & 
\end{flalign}
\begin{flalign}
& s+X_{\alpha} \to \overline{u}+\overline{b} &  \nonumber \\
& E_{a}E_{b}\sigma v  = \frac{3 |\kappa_{\alpha}|^{2} p_{f} }{2  \pi\sqrt{\hat{s}} } \left[ E_{a}E_{b}E_{c}E_{d} + \frac{1}{3}p_{i}^{2}p_{f}^{2} \right] &
\end{flalign}
\begin{flalign}
& X_{2} + \overline{u} \to X_{1}+\overline{u} & \nonumber \\
& E_{a}E_{b}\sigma v  = \frac{3 |\kappa_{3}|^{2} p_{f} }{ 8 \pi\sqrt{\hat{s}} } \left[ 2E_{a}E_{b}E_{c}E_{d} + E_{c}E_{d}p_{i}^{2} + E_{a}E_{b}p_{f}^{2} + \frac{4}{3}p_{i}^{2}p_{f}^{2} + \frac{ \mathrm{Re}[\kappa_{3}\kappa_{3}] }{ |\kappa_{3}|^{2} }M_{X2}M_{X1}E_{b}E_{d} \right] &
\end{flalign}
\begin{flalign}
& X_{2}+X_{1} \to \overline{u}u &  \nonumber \\
& E_{a}E_{b}\sigma v  = \frac{3 |\kappa_{3}|^{2} p_{f} }{ 16 \pi\sqrt{\hat{s}} } \left[ 2E_{a}E_{b}E_{c}E_{d}  + \frac{2}{3}p_{i}^{2}p_{f}^{2} - \frac{ \mathrm{Re}[\kappa_{3}\kappa_{3}] }{ |\kappa_{3}|^{2} }M_{X2}M_{X1}(E_{c}E_{d}+p_{f}^{2}) \right] &
\end{flalign}
\begin{flalign}
& X_{\alpha}+X_{\alpha} \to \overline{u}u  & \nonumber \\
& E_{a}E_{b}\sigma v  = \frac{3 |\kappa_{\alpha+3}|^{2} p_{f} }{ 16 \pi\sqrt{\hat{s}} } \left[ 2E_{a}E_{b}E_{c}E_{d}  + \frac{2}{3}p_{i}^{2}p_{f}^{2} - M_{X\alpha }^{2}(E_{c}E_{d}+p_{f}^{2}) \right]. &
\end{flalign}

\tocless\subsection{Cross sections --- CP violation}
\underline{CP violation for $u_{R}+X_{1}\to u_{R}+X_{2}$:}\\
This CP violation arises from the interference of two tree level diagrams (corresponding to different continuous fermion lines) with a loop level diagram.
\begin{align}
 E_{a}E_{b}(\sigma-\overline{\sigma}) v  =\frac{3p_{f}}{ \pi \sqrt{\hat{s}}} g_{1}(\hat{s},m_{s},m_{b}) \Bigg \{ &  2\mathrm{Im}[\kappa_{3}\kappa_{1}\kappa_{2}^{\ast}]\Big[E_{a}E_{b}+p_{i}^{2}\Big]\Big[E_{c}E_{d}+p_{f}^{2}\Big] \nonumber \\
& \quad + \mathrm{Im}[\kappa_{3}^{\ast}\kappa_{1}\kappa_{2}^{\ast}]M_{X1}M_{X2}E_{b}E_{d} \Bigg \}, 
\end{align}
where:
\begin{equation}
g_{1}(\hat{s},m_{i},m_{j})=\frac{1}{32\pi} \Bigg[\hat{s}-(m_{i}^{2}+m_{j}^{2})\Bigg]\sqrt{1-\frac{2(m_{i}^{2}+m_{j}^{2})}{\hat{s}}+\frac{(m_{i}^{2}-m_{j}^{2})^{2}}{\hat{s}^{2}}  }.
\end{equation}
\underline{CP violation for $u_{R}+X_{1L}\to \overline{s}+\overline{b}$:} \\
This CP violation arises from the interference of a tree level diagram with two different loop level diagrams.
\begin{align}
 E_{a}E_{b}(\sigma-\overline{\sigma}) v & =  \frac{6p_{f}}{\pi \sqrt{\hat{s}}}\mathrm{Im}[\kappa_{1}^{\ast}\kappa_{2}\kappa_{3}^{\ast}]g_{1}(\hat{s},m_{u},M_{X2})\Big[E_{a}E_{b}+p_{i}^{2}\Big]\Big[E_{c}E_{d}+p_{f}^{2}\Big] \nonumber \\
&  \quad -\frac{3p_{f}}{\pi \sqrt{\hat{s}}}\mathrm{Im}[\kappa_{1}^{\ast}\kappa_{2}\kappa_{3}]E_{b}g_{2}(\hat{s},m_{u},M_{X2})M_{X1}M_{X2}\Big[E_{c}E_{d}+p_{f}^{2}\Big],
\end{align}
where: 
\begin{equation}
g_{2}(\hat{s},m_{i},m_{j})=-\frac{1}{32\pi\sqrt{\hat{s}}}\Bigg[\hat{s}+m_{i}^{2}-m_{j}^2\Bigg]\sqrt{1-\frac{2(m_{i}^{2}+m_{j}^2)}{\hat{s}}+\frac{(m_{i}^{2}-m_{j}^2)^{2}}{\hat{s}^{2}}}.
\end{equation}
\underline{CP violation for $u_{R}+X_{2}\to \overline{s_{R}}+\overline{b_{R}}$:}\\
This CP violation arises from the interference of a tree level diagram with two different loop level diagrams.
\begin{align}
 E_{a}E_{b}(\sigma-\overline{\sigma}) v &  = \frac{6p_{f}}{\pi \sqrt{\hat{s}}}\mathrm{Im}[\kappa_{2}^{\ast}\kappa_{1}\kappa_{3}]g_{1}(\hat{s},m_{u},M_{X1})\Big[E_{a}E_{b}+p_{i}^{2}\Big]\Big[E_{c}E_{d}+p_{f}^{2}\Big] \nonumber \\
 & \quad  -\frac{3p_{f}}{\pi \sqrt{\hat{s}}}\mathrm{Im}[\kappa_{2}^{\ast}\kappa_{1}\kappa_{3}^{\ast}]E_{b}g_{2}(\hat{s},m_{u},M_{X1})M_{X1}M_{X2}\Big[E_{c}E_{d}+p_{f}^{2}\Big].
\end{align}

\tocless\subsection{Decay rates \label{sec:appdec}}
The decay rates which appear in the Boltzmann equations are thermally averaged~\cite{Davidson:2008bu}:
	\begin{equation}
	\Gamma^{\rm th} = \frac{K_{1}(M/T)}{K_{2}(M/T)}\Gamma,
	\end{equation}
where $M$ is the mass of the decaying particle, $\Gamma$ is the decay rate in the rest frame of the decaying particle and $K_{n}(x)$ is the modified Bessel function of the second kind of order $n$.
\subsubsection{\underline{$\Gamma(X_{1L} \to usb)$}}
Ignoring the final state masses the decay width is:
\begin{equation}
\frac{\Gamma_{1A}}{2} = \frac{|\kappa_{1}|^{2}M_{X1}^{5}}{1024\pi^{3}},
\end{equation}
so $\Gamma_{1A}$ is the sum over the partial widths $\Gamma(X_{1L} \to usb)+\Gamma(X_{1L} \to \overline{usb})$.

\subsubsection{\underline{$\Gamma(X_{2} \to usb)$}}
Ignoring the final state masses the decay width is:
\begin{equation}
\frac{\Gamma_{2A}}{2} = \frac{|\kappa_{2}|^{2}M_{X2}^{5}}{1024\pi^{3}},
\end{equation}
so $\Gamma_{2A}$ is the sum over the partial widths $\Gamma(X_{2} \to usb)+\Gamma(X_{2} \to \overline{usb})$.

\subsubsection{\underline{$\Gamma(X_{2} \to X_{1L}u\overline{u})$}}
For this decay we take into account the mass of $X_{1}$. In integral form the width is given by:
	\begin{align}
	\Gamma_{X2B}=\frac{3|\kappa_{3}|^{2}M_{X2}^{5}}{512\pi^{3}} \int dx_{a} \int dx_{b} \bigg \{ &
	x_{a}(1+\mu_{12}-x_{a})+x_{b}(1-\mu_{12}-x_{b}) \nonumber \\
	&\quad +\frac{2\mathrm{Re}[\kappa_{3}^2]}{|\kappa_{3}|^{2}}\frac{M_{X1}}{M_{X2}}(1-\mu_{12}-x_{b}) \bigg \},
	\end{align}
where $\mu_{12}\equiv (M_{X1}/M_{X2})^{2}$ and the limits of integration are:
	\begin{equation}
	\frac{1}{2}\left([2 - x_{a}] - \sqrt{x_{a}^{2} - 4\mu_{12}}\right)  \leq  x_{b}  \leq  \frac{1}{2}\left([2 - x_{a}] + \sqrt{x_{a}^{2} - 4\mu_{12}}\right),
	\end{equation}
	\begin{equation}	
	2\sqrt{\mu_{12}}  \leq  x_{a}  \leq  1 + \mu_{12}. 
	\end{equation}	

\tocless\subsection{CP violation in the $X_{2}$ decay}
This CP violation arises from the interference of a tree level diagram with two different loop level diagrams.
The tree level amplitude for $X_{2} \to usb$ with momenta $p_{i}$, $p_{a}$, $p_{b}$, $p_{c}$ respectively is given by:
\begin{equation}
\mathcal{M}_{T}=-i\kappa_{2}^{\ast}\Big[\overline{u}(p_{a})\mathrm{L}u(p_{i})\Big]\Big[\overline{u}(p_{c})\mathrm{L}v(p_{b})\Big].
\end{equation}
Here L (R) denotes the left (right) projection operator. The first loop diagram has amplitude:
\begin{equation}
\mathcal{M}_{L1}=\kappa_{3}^{\ast}\kappa_{1}^{\ast}\Big[\overline{u}(p_{a})Lu(p_{i})\Big]\Big[\overline{u}(p_{c})\mathrm{L}v(p_{b})\Big]\int \frac{d^{4}k}{(2\pi)^{4}}\frac{\mathrm{Tr}\Big[\mathrm{R}(\slashed{p_{t}}+\slashed{k})\slashed{k}\Big]}{\Big([p_{t}+k]^{2}-M_{X1}^{2}\Big)\Big(k^{2}-m_{u}^{2}\Big)},
\end{equation}
where $p_{t}=p_{i}-p_{a}$. The second loop diagram has amplitude:
\begin{equation}
\mathcal{M}_{L2}=\kappa_{1}^{\ast}\kappa_{3}M_{X1}\Big[\overline{u}(p_{c})\mathrm{L}v(p_{b})\Big]\int \frac{d^{4}k}{(2\pi)^{4}}\frac{\Big[\overline{v}(p_{i})\mathrm{R}\slashed{k}v(p_{a})\Big]}{\Big([p_{t}+k]^{2}-M_{X1}^{2}\Big)\Big(k^{2}-m_{u}^{2}\Big)}.
\end{equation}
The total CP violation can be expressed as:
	\begin{equation}
	\epsilon_{D}=\gamma_{1}+\gamma_{2},
	\end{equation} 
where $\gamma_{1}$ ($\gamma_{2}$) comes from the interference between $\mathcal{M}_{T}^{\ast}$ and $\mathcal{M}_{L1}$ ($\mathcal{M}_{L2}$).

\subsubsection{Contribution from $\mathcal{M}_{T}^{\ast}\mathcal{M}_{L1}$}
\begin{align}
\gamma_{1} \Gamma_{2A} = 3\mathrm{Im}[\kappa_{2}\kappa_{1}^{\ast}\kappa_{3}^{\ast}]\frac{M_{X2}^{5}}{512 \pi^{4}} \int_{0}^{1-\mu_{12}}dx_{a} x_{a}^{2}(1-x_{a})g_{3}\left(\frac{M_{X2}x_{a}}{2}\right),
\end{align}
where,
\begin{align}
g_{3}(E)=\int_{-1}^{1}dc_{\theta} \frac{ x_{0}(x_{0}^{2}+x_{0}\sqrt{ x_{0}^{2}+E^{2}-2x_{0}Ec_{\theta}+M_{X1}^{2} }-x_{0}Ec_{\theta}) }{ \sqrt{ x_0^{2}+E^{2}-2x_{0}Ec_{\theta}+M_{X1}^{2} }+x_{0}-Ec_{\theta} }
\end{align}
and
\begin{align}
\label{eq:xnought}
x_{0}=\frac{(M_{X2}^{2}-2M_{X2}E-M_{X1}^{2})}{2(M_{X2}-E-Ec_{\theta})}.
\end{align}

\subsubsection{Contribution from  $\mathcal{M}_{T}^{\ast}\mathcal{M}_{L2}$}

\begin{align}
\gamma_{2} \Gamma_{2A} = 3\mathrm{Im}[\kappa_{2}\kappa_{1}^{\ast}\kappa_{3}]\frac{ M_{X1}M_{X2}^{4} }{ 512\pi^{4} }\int_{0}^{1-\mu_{12}} dx_{a} x_{a}(1-x_{a})g_{4}\left(\frac{M_{X2}x_{a}}{2}\right),
\end{align}
where,
\begin{align}
g_{4}(E)= \int_{-1}^{1}dc_{\theta} \frac{x_{0}^{2}E(1+c_{\theta})}{(x_{0}^{2}-2x_{0}Ec_{\theta}+E^{2}+M_{X1}^{2})^{1/2}+x_{0}-Ec_{\theta} },
\end{align}
and the expression for $x_{0}$ is given in eq. (\ref{eq:xnought}).

\bibliographystyle{jhep}
\providecommand{\href}[2]{#2}\begingroup\raggedright\endgroup

\end{document}